%% file: neurips_2026.tex
\documentclass{article}

\PassOptionsToPackage{numbers, compress}{natbib}
\usepackage[preprint]{neurips_2026}

\usepackage[utf8]{inputenc} 
\usepackage[T1]{fontenc}    
\usepackage{hyperref}       
\usepackage{url}            
\usepackage{booktabs}       
\usepackage{amsfonts}       
\usepackage{nicefrac}       
\usepackage{microtype}      
\usepackage{xcolor}         
\usepackage{graphicx}
\usepackage{amsmath}
\usepackage{subcaption}
\usepackage{caption}
\usepackage{float}
\usepackage{wrapfig}
\usepackage{enumitem}
\usepackage{colortbl}
\usepackage{adjustbox}

\title{PersonaGest: Personalized Co-Speech Gesture Generation with Semantic-Guided Hierarchical Motion Representation}

\author{%
  Junchuan Zhao$^{*}$ \quad Qifan Liang$^{*}$ \quad Ye Wang \\[0.4em]
  School of Computing, National University of Singapore \\
  \texttt{\{junchuan, liangqifan\}@u.nus.edu, dcswangy@nus.edu.sg}
  \thanks{Equal contribution.}
}

\usepackage{titletoc}

\titlecontents{section}
  [1.8em]
  {\bfseries}
  {\contentslabel{1.8em}}
  {}
  {\titlerule*[0.5pc]{.}\contentspage}

\titlecontents{subsection}
  [4.2em]
  {}
  {\contentslabel{2.8em}}
  {}
  {\titlerule*[0.5pc]{.}\contentspage}

\begin{document}

\maketitle

\begin{abstract}
\input{sections/abstract}
\end{abstract}

\section{Introduction}\label{sec:intro}
\input{sections/intro}

\section{Related Works}\label{sec:related}
\input{sections/rela}

\section{Methodology}\label{sec:method}
\input{sections/method}

\section{Experiments}\label{sec:exp_setting}
\input{sections/exp}

\section{Results}\label{sec:result}
\input{sections/results}

\section{Conclusion}\label{sec:conclusion}
\input{sections/concl}

\bibliographystyle{plainnat}
\bibliography{reference}

\clearpage
\appendix
\input{sections/appendix}



\end{document}

%% file: sections/abstract.tex
Co-speech gesture generation aims to synthesize realistic body movements that are semantically coherent with speech and faithful to a user-specified gestural style. Existing VQ-VAE based co-speech gesture generation methods improve generation quality but fail to encode semantic structure into the motion representation or explicitly disentangle content from style, limiting both semantic coherence and personalization fidelity. We present PersonaGest, a two-stage framework addressing both limitations. In the first stage, a semantic-guided RVQ-VAE disentangles motion content and gestural style within the residual quantization structure, where a Semantic-Aware Motion Codebook (SMoC) organizes the content codebook by gesture semantics and contrastive learning further enforces content-style separation. In the second stage, a Masked Generative Transformer generates content tokens via a semantic-aware re-masking strategy, followed by a cascade of Style Residual Transformers conditioned on a reference motion prompt for style control. Extensive experiments demonstrate state-of-the-art performance on objective metrics and perceptual user studies, with strong style consistency to the reference prompt. Our project page with demo videos is available at \url{https://danny-nus.github.io/PersonaGest/}.

%% file: sections/intro.tex
Co-speech gesture generation aims to synthesize realistic body movements that are synchronized and semantically aligned with speech \cite{nyatsanga2023comprehensive}. Beyond words, human gestures carry rich communicative intent through motion cues that speech alone cannot fully capture \citep{mcneill1992hand, tipper2015body}. Moreover, gesture is inherently personal: each speaker exhibits a characteristic style that persists across different linguistic content and contexts. This motivates personalized co-speech gesture generation for digital humans and virtual avatars \citep{lee2024all, van1998persona, ali2025expanding, dang2025user}, which aims to produce gestures that are both speech-coherent and faithful to a user-specified style.

\begin{figure*}[t]
  \centering
  \includegraphics[width=\textwidth]{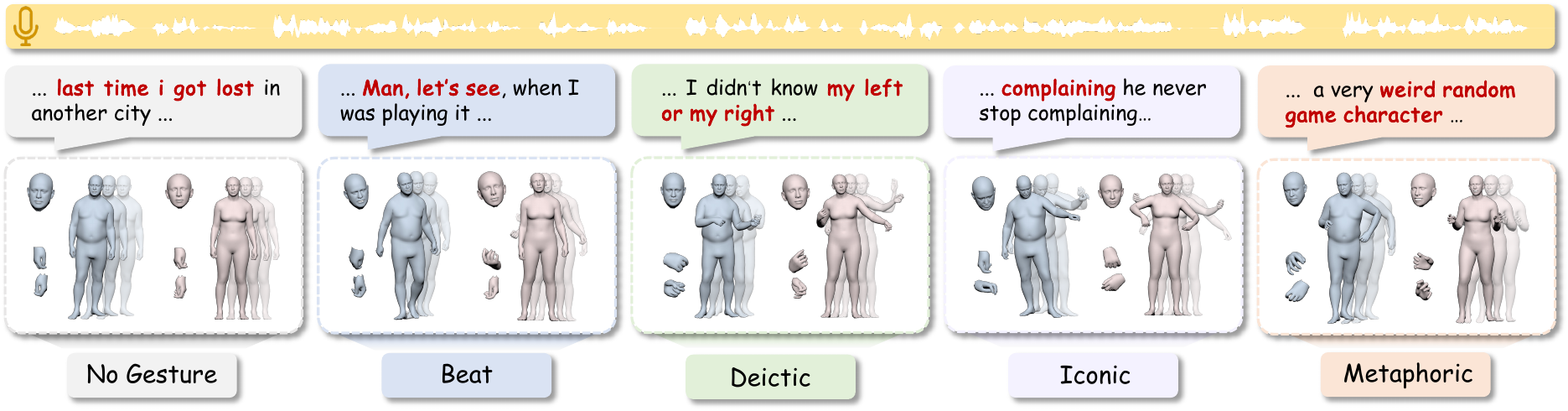}
  \caption{Co-speech gestures exhibit diverse semantic types, including beat, deictic, iconic, and metaphoric gestures, that vary across speakers and speech content, motivating the need for style-controllable gesture generation.}
  \label{fig:motivation}
  \vspace{-5mm}
\end{figure*}

Co-speech gesture generation has witnessed substantial progress in recent years. VQ-VAE-based approaches have emerged as the dominant paradigm, encoding continuous motion into discrete latent codes as compact motion priors, and leveraging generative models to predict these codes conditioned on speech~\citep{yinpyramotion, liu2025semges, zhang2025semtalk, chen2025motion, yi2023generating, guo2024momask, liu2025gesturelsm, voss2023aq}. Despite these advances, existing methods predominantly treat gesture generation as a speaker-agnostic audio-to-motion mapping problem, without explicitly accounting for the personal and semantic dimensions of human gesture, as illustrated in Figure~\ref{fig:motivation}.

Personalized gesture generation remains underexplored. Example-based methods~\citep{chen2024enabling, ghorbani2023zeroeggs, liu2025mimicparts} offer greater flexibility than label-based approaches~\citep{ahuja2020style, yang2023diffusestylegesture} by using reference motion clips as implicit style signals. However, how to effectively disentangle motion content and gestural style in the latent space remains an open problem. In existing VQ-VAE-based co-speech gesture generation frameworks, content and style are entangled within the same discrete representation, making it difficult to transfer a user-specified style to novel speech inputs without distorting the semantic structure of the generated gestures.

A separate challenge concerns gesture semantics. While recent works incorporate semantic labels into the generation process~\citep{zhi2023livelyspeaker, mughal2024convofusion, zhang2025semtalk, liu2025semges}, they treat semantics as a supervisory signal applied on top of an already-learned representation, without encoding semantic structure into the motion space itself. As a result, each motion token carries no inherent semantic meaning, limiting how precisely the generated gestures can reflect the communicative intent of the speech.

To address these challenges, we propose \textbf{PersonaGest}, a two-stage framework for personalized and semantically coherent co-speech gesture generation. In the first stage, a semantic-guided RVQ-VAE disentangles motion content and gestural style within the residual quantization structure, where the first residual layer forms a semantically-organized content codebook supervised by gesture category labels and the remaining layers encode gestural style. In the second stage, a Masked Generative Transformer generates content tokens via a semantic-aware re-masking strategy, followed by a cascade of Style Residual Transformers that generates style tokens layer by layer conditioned on a reference motion prompt. 

Our contributions are summarized as follows:
\vspace{-2mm}
\begin{itemize}[leftmargin=*, itemsep=1pt, topsep=2pt, parsep=0pt]
\item We propose a semantic-guided RVQ-VAE that disentangles motion content and gestural style within the residual quantization structure via contrastive learning, with a Semantic-Aware Motion Codebook (SMoC) that partitions the content codebook into semantic regions and routes quantization by predicted gesture category.
\item We introduce a semantic-aware re-masking strategy for the Masked Generative Transformer, where predicted semantic confidence scores dynamically modulate the re-masking probability during iterative refinement, yielding content tokens with improved semantic coherence. 
\item Extensive experiments demonstrate state-of-the-art performance on objective metrics and perceptual user studies, with strong style consistency between generated motion and reference motion prompt.
\end{itemize}

%% file: sections/rela.tex
\subsection{Co-Speech Gesture Generation}
\label{rel:co-speech}

Co-speech gesture generation has progressed from rule-based systems \citep{cassell1994animated, kopp2006towards} to data-driven models based on RNNs \citep{ginosar2019learning, qian2021speech}, transformers \citep{qi2024emotiongesture, bhattacharya2021text2gestures}, and diffusion models \citep{chen2024diffsheg, yang2023diffusestylegesture, zhang2026mitigating, mughal2024convofusion}. Recent methods increasingly follow a VQ-VAE-based two-stage paradigm, first learning a discrete motion codebook and then predicting speech-conditioned motion tokens with autoregressive models \citep{liu2022audio, yi2023generating, xiao2024eggesture}, masked-token refinement \citep{liu2024emage, guo2024momask, liu2025semges, zhang2025echomask, mao2024mdt}, latent diffusion \citep{ji2023c2g2}, flow matching \citep{liu2025gesturelsm}, or LLM-based motion reasoning \citep{chen2025motion}. To improve gesture semantics, prior works introduce CLIP-guided script planning \citep{zhi2023livelyspeaker}, LLM-based semantic gesture retrieval and fusion \citep{zhang2024semantic}, rhythmic-semantic motion gating \citep{zhang2025semtalk}, or explicit semantic alignment losses \citep{liu2025semges}. However, these methods mainly impose semantic constraints during generation, leaving the underlying motion representation without explicit semantic organization. This motivates codebook-level semantic modeling, where semantic structure is embedded directly into the motion codebook to provide semantically meaningful content tokens.

\subsection{Personalized Gesture Generation}
\label{rel:per}
Personalized gesture generation aims to synthesize gestures that reflect individual gestural styles. Early methods approached this through label-based style control, conditioning generation on discrete style descriptors~\citep{ahuja2020style, wu2021modeling, habibie2022motion}, but predefined labels are inherently coarse and struggle to capture fine-grained gestural variation across individuals. Another line of work explored speaker adaptation, adjusting pretrained gesture models to target speakers using limited data~\citep{ahuja2022low}, later extended to continual multi-speaker settings~\citep{ahuja2023continual}. More recent approaches have shifted toward example-based and prompt-based style control, which offer greater flexibility without requiring predefined descriptors, including zero-shot style encoding from short motion clips~\citep{ghorbani2023zeroeggs}, text prompt-guided generation through joint speech-text-motion embeddings~\citep{chen2024enabling}, localized body-part style modeling~\citep{liu2025mimicparts}, and LLM-based interpretation of diverse style references~\citep{chen2025motion}. Despite these advances, existing methods mainly treat style as an external conditioning signal during generation, while the interaction between style control and gesture semantics remains underexplored.  Most related to our work, VQ-Style~\citep{zargarbashi2026vq} disentangles content and style within 
the RVQ-VAE hierarchy for general motion sequences, but does not incorporate 
gesture semantics into the discrete representation, a gap that PersonaGest 
directly addresses.

%% file: sections/method.tex


PersonaGest consists of two stages. In the first stage, a RVQ-VAE learns disentangled motion representations, where the first codebook captures semantic-aware motion content via SMoC and the remaining codebooks encode gestural style. In the second stage, following MoMask \citep{guo2024momask}, a Content Masked Transformer generates content tokens from speech audio and speaker identity via semantic-aware re-masking, while a cascade of Style Residual Transformers predicts style tokens conditioned on the generated content tokens and a reference motion prompt, followed by motion reconstruction through the RVQ-VAE decoder.

\subsection{Semantic-Aware Motion Representation Learning}
\label{method:smoc}
\paragraph{Overall Framework.}

\begin{figure}
\centering
\includegraphics[width=1.0\textwidth]{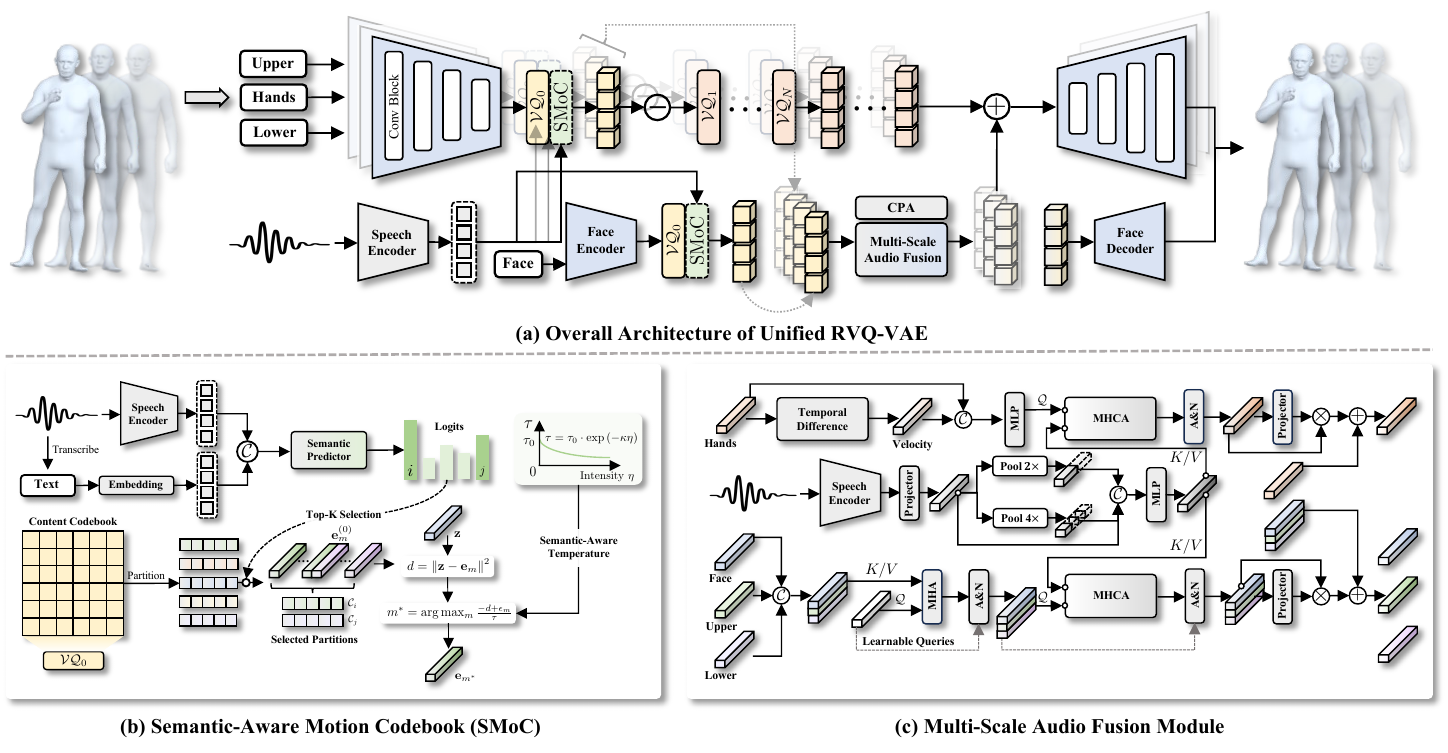}
\caption{Overall architecture of the semantic-aware RVQ-VAE. The content codebook $\mathcal{C}^p$ is organized by 
gesture semantics (SMoC), while the residual layers 
capture gestural style. Multi-Scale Audio Fusion (MSAF) module injects multi-scale audio information into the content latents.\label{fig:stage1}}
\vspace{-3mm}
\end{figure}

We represent full-body motion as four body-part sequences 
$\mathbf{m}^p \in \mathbb{R}^{T \times D_p}$ for 
$p \in \mathcal{B} = \{\text{upper, hands, lower, face}\}$, 
where $T$ denotes the sequence length and $D_p$ the part-specific dimension. 
Each part is encoded by $\mathcal{E}_p(\cdot)$ into latent representations 
$\mathbf{z}^p \in \mathbb{R}^{T' \times D}$, where $T' = T/4$ due to temporal downsampling. 
For body parts $p \in \{\text{upper, hands, lower}\}$, 
$\mathbf{z}^p$ is quantized through residual vector quantization, 
where the first codebook $\mathcal{C}^p$ models semantic motion content under SMoC supervision, while the remaining residual codebooks 
$\{\mathcal{S}_n^p\}_{n=1}^{N}$ capture gestural style. 
For the face branch, we employ a standard VQ-VAE without explicit content-style disentanglement, as facial motion is primarily determined by speech-driven lip synchronization, whereas gestural style is mainly expressed through body and hand movements~\citep{mcneill1992hand}. 
The quantized content latents from all body parts, together with speech audio $\mathbf{a}$, are fused through a Multi-Scale Audio Fusion (MSAF) module and a Cross-Part Attention (CPA) module, before being decoded by part-specific decoders $\mathcal{D}_p(\cdot)$ to reconstruct the original motion sequences. 
The overall pipeline of our motion RVQ-VAE is depicted in Figure \ref{fig:stage1} and summarized as follows:
\begin{equation}
\begin{aligned}
\mathbf{z}^p &= \mathcal{E}_p(\mathbf{m}^p), \quad
\mathbf{z}^p_c, \mathbf{z}^p_s = \mathrm{RVQ}(\mathbf{z}^p), \\
\hat{\mathbf{z}}^p_c &= \mathrm{CPA}(\mathrm{MSAF}(\{\mathbf{z}^p_c\}_{p \in \mathcal{B}}, \mathbf{a})), \quad
\hat{\mathbf{m}}^p = \mathcal{D}_p(\hat{\mathbf{z}}^p_c + \mathbf{z}^p_s),
\end{aligned}
\end{equation}
where $\mathcal{B} = \{\text{upper, hands, lower, face}\}$. 
For the face part, a plain VQ is used without style quantization, and $\mathbf{z}^{\text{face}}_s = \mathbf{0}$.

\paragraph{Semantic-Aware Motion Codebook (SMoC).}

Inspired by semantic-guided codebook organization~\citep{ding2024sgc} and sparse top-$K$ routing in mixture-of-experts~\citep{shazeer2017outrageously}, SMoC partitions the content codebook $\mathcal{C}^p$ into $K$ semantic regions $\{\mathcal{C}^p_k\}_{k=1}^{K}$, each corresponding to a gesture category defined in~\citep{liu2024emage} (e.g., iconic, metaphoric, deictic). Given raw audio $\mathbf{a}$ and text transcript $\mathbf{w}$, a speech encoder and text embedding produce features $\mathbf{e}_a, \mathbf{e}_w \in \mathbb{R}^{T' \times D}$, which are concatenated and passed to a semantic predictor $\psi(\cdot)$ to obtain per-frame semantic logits $\mathbf{q}$. The top-$K$ predicted categories determine the active partitions $\mathcal{C}^p_{k_i}$ and $\mathcal{C}^p_{k_j}$ from the content codebook $\mathcal{C}^p$, and the nearest codebook entry is retrieved from their union. 
During training, Gumbel sampling with a semantic-aware 
temperature is applied to encourage codebook exploration:
\begin{equation}
    m^* = \arg\max_{m \in \mathcal{C}^p_{k_i} \cup \mathcal{C}^p_{k_j}} 
    \left(\frac{-\|\mathbf{z}^p - \mathbf{e}_m\|^2}{\tau} + g_m\right), 
\end{equation}
where $\mathbf{e}_m$ is the codebook embedding and $g_m \sim \text{Gumbel}(0,1)$ is a Gumbel noise sample.
The temperature $\tau = \tau_0 \cdot \exp(-\kappa \cdot \eta)$ is modulated by the gesture intensity $\eta \in [0,1]$ provided by training data annotations, encouraging sharper assignments for high-intensity frames during training. At inference, Gumbel noise is removed and the selection reduces to deterministic nearest-neighbor search. After first layer content quantization, we obtain the content quantized latent $\mathbf{z}^p_c = \mathbf{e}_{m^*}$.

\paragraph{Multi-Scale Audio Fusion (MSAF).}
MSAF injects multi-scale audio information into each body part's content latent through two parallel branches. For hands, whose dynamics are more informative than absolute positions due to rapid fine-grained articulations, a velocity feature is computed via temporal difference and projected to form the motion query $\mathbf{Q}^{\text{hands}}$. For remaining parts $p \in \{\text{upper, lower, face}\}$, a shared learnable global token $\mathbf{Q}_{\text{learn}}$ queries stacked part features through multi-head attention (MHA) to capture cross-part context, and is added and normalized with each part's feature to form part-specific queries $\mathbf{Q}^p$. For the audio branch, $\mathbf{e}_a$ is average-pooled at $2\times$ and $4\times$ temporal scales, concatenated with $\mathbf{e}_a$, and projected to form multi-scale keys $\mathbf{K}_a$ and values $\mathbf{V}_a$. 
Each part then attends to the audio features via multi-head cross attention (MHCA):
\begin{equation}
\tilde{\mathbf{z}}^p_c =
\mathrm{MHCA}(\mathbf{Q}^p,\mathbf{K}_a,\mathbf{V}_a),\;
\tilde{\mathbf{z}}^{\rm hands}_c =
\mathrm{MHCA}(\mathbf{Q}^{\rm hands},\mathbf{K}_a,\mathbf{V}_a), 
\end{equation}
where $p \in \{\text{upper, lower, face}\}$. 
Finally, the MHCA output is fused with the original latent via a gated residual connection for all four parts to obtain audio-enriched content representations:
\begin{equation}
    \mathbf{z}^p_{c} \leftarrow \mathbf{z}^p_{c} + 
    \sigma\!\left(\mathbf{W}_g \tilde{\mathbf{z}}^p_c\right) 
    \odot \tilde{\mathbf{z}}^p_c
\end{equation}
where $\sigma(\cdot)$ is the sigmoid function and 
$\mathbf{W}_g$ is a learned linear projection.

\paragraph{Cross-Part Attention (CPA).} 
CPA enables inter-part communication by treating each body 
part as a token at each time step. The content latents from MSAF are stacked into $\mathbf{h}_c \in \mathbb{R}^{T' \times 4 \times D}$, where learnable positional embeddings are added along the part dimension, and processed by standard Transformer blocks with 
pre-norm multi-head self-attention (MHSA) and feed-forward 
network (FFN). The output is split back into 
per-part latents $\{\mathbf{z}^p_c\}_{p\in\mathcal{B}}$ for subsequent decoding.

\paragraph{Training Objectives.}
The RVQ-VAE is trained with a combination of reconstruction, 
commitment, and disentanglement losses. The reconstruction 
loss $\mathcal{L}_{\text{rec}}$ is computed as Geodesic 
loss~\citep{tykkala2011direct} between the decoded and ground-truth motion 
for all body parts, with an additional velocity loss 
$\mathcal{L}_{\rm{vel}}$ computed as L1 loss on 
first-order finite differences to improve temporal smoothness.

To enforce content-style disentanglement, inspired by 
VQ-Style~\citep{zargarbashi2026vq} and 
MimicParts~\citep{liu2025mimicparts}, we apply a contrastive 
loss $\mathcal{L}_{\text{cl}}$ on the style latent space using an 
InfoNCE objective, where motion clips from the same speaker 
form positive pairs and clips from different speakers serve 
as negatives:
\begin{equation}
    \mathcal{L}_{\text{cl}} = -\frac{1}{|\mathcal{S}^+|}\sum_{(i,j)\in\mathcal{S}^+}
\log\frac{\exp\!\left(\text{cos\_sim}(\mathbf{z}^i_s,\,\mathbf{z}^j_s)/\tau_{\text{cl}}\right)}
{\sum_{k\neq i}\exp\!\left(\text{cos\_sim}(\mathbf{z}^i_s,\,\mathbf{z}^k_s)/\tau_{\text{cl}}\right)}, 
\end{equation}
where $\mathcal{S}^+$ denotes the set of same-speaker 
positive pairs in the batch. Inspired by NaturalSpeech3~\citep{ju2024naturalspeech}, 
which employs phoneme supervision to constrain the content 
codec in FACodec, we apply a phoneme prediction loss 
$\mathcal{L}_{\text{phone}}$ on the content latent via a phoneme predictor $\varphi(\cdot)$:
\begin{equation}
    \mathcal{L}_{\text{phone}} = 
    \mathrm{CE}\!\left(\varphi(\mathbf{z}^p_c),\; 
    \mathbf{y}_{\text{phone}}\right).
\end{equation}
The semantic predictor $\psi(\cdot)$ is supervised with a 
multi-label classification loss $\mathcal{L}_{\text{sem}}$. The total training objective is:
\begin{equation}
\begin{aligned}
    \mathcal{L}_{\text{RVQ-VAE}} = \,
    &\mathcal{L}_{\text{rec}} + 
    \mathcal{L}_{\text{vel}} + 
    \mathcal{L}_{\text{sem}} + 
    \mathcal{L}_{\text{cl}} + 
    \mathcal{L}_{\text{phone}} \\
    & +\, \sum_{n=0}^{N} \left( 
    \bigl\|{\rm sg}(\mathbf{z}^p_n) - \mathbf{e}^p_{n}\bigr\|^2_2 + 
    \bigl\|\mathbf{z}^p_n - {\rm sg}(\mathbf{e}^p_{n})\bigr\|^2_2 
    \right), 
\end{aligned}
\end{equation}
where $\mathbf{z}^p_n$ and $\mathbf{e}^p_n$ denote the residual and nearest 
codebook entry at layer $n$, 
and ${\rm sg}(\cdot)$ denotes stop-gradient. 

\subsection{Semantic-Aware Personalized Gesture Generation}
\label{method:stage2}
\begin{figure}
\centering
\includegraphics[width=1.0\textwidth]{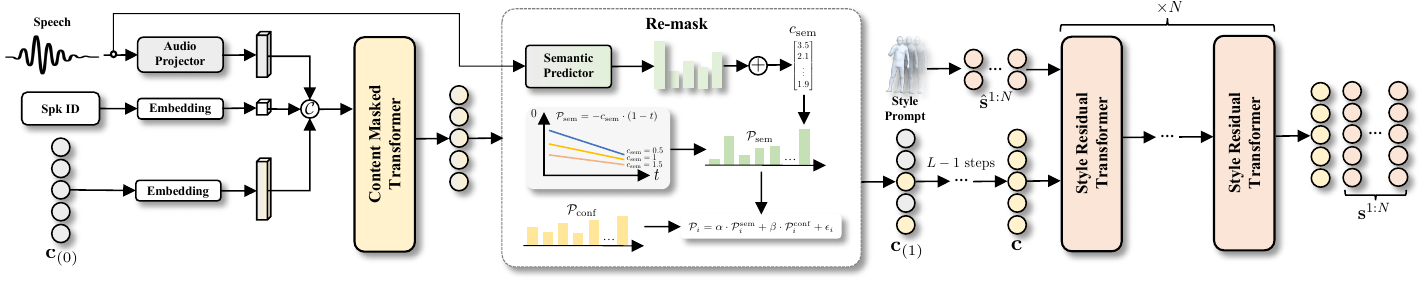}
\caption{Two-stage non-autoregressive gesture token generation. 
A Content Masked Transformer generates content tokens via semantic-aware re-masking, followed by a cascade of $N$ Style Residual Transformers that generates style tokens conditioned 
on a reference motion prompt.\label{fig:stage2}}
\vspace{-3mm}
\end{figure}

As illustrated in Figure~\ref{fig:stage2}, the generation 
pipeline consists of two stages. In the first stage, a 
Content Masked Transformer (CMT) generates content tokens 
$\mathbf{c}$ conditioned on speech and speaker identity. 
In the second stage, a cascade of Style Residual Transformers 
(SRT) generates style tokens $\mathbf{s}^{1:N}$ conditioned 
on the predicted content tokens and a reference motion prompt. 
An individual CMT and SRT are trained for each body part, 
except for the face which is handled by a CMT only. 

\subsubsection{Content Masked Transformer}
\label{method:stage2-1}
Given the discrete motion tokens from Stage 1, we train a 
Content Masked Transformer (CMT) to generate content tokens 
$\mathbf{c}$ conditioned on speech and speaker identity, 
following the masked generative modeling framework of 
MoMask~\citep{guo2024momask}. The speaker identity and speech first are encoded into 
embeddings $\mathbf{e}_{\text{spk}}$ and $\mathbf{e}_a$ 
respectively. During training, a subset of motion tokens 
are replaced with mask token \texttt{[MASK]} to form $\tilde{\mathbf{c}}$, 
which is then embedded and fed into the Transformer prepended with 
$\mathbf{e}_c = [\mathbf{e}_{\text{spk}}; \mathbf{e}_a]$ as conditioning signals.

\paragraph{Semantic-Aware Masking and Remasking.}
We propose a semantic-aware masking strategy that prioritizes tokens by predicted semantic class, inspired by~\citep{kong2023priority} where more informative tokens should be established earlier to guide subsequent generation. Given semantic logits $\mathbf{q}_i$ from $\psi(\cdot)$, we compute a normalized semantic score $\bar{q}_{i} \in [0,1]$ per token. The masking priority at timestep $t \in [0,1]$ is defined as $\mathcal{P}^{\rm sem}_i = -\bar{q}_{i} \cdot (1 - t)$, so that semantically ambiguous tokens (low $\bar{q}$) are masked first, with priority decaying to zero as $t \rightarrow 1$ to recover uniform random selection. A randomisation scale $r$ further controls the proportion of tokens masked uniformly at random rather than by priority. The masking ratio follows a cosine schedule~\citep{chang2022maskgit} $\gamma(t) = \cos\!\left(\frac{\pi t}{2}\right) \in [0,1]$, and the training objective minimizes the negative log-likelihood over masked positions:
\begin{equation}
    \mathcal{L}_{\text{mask}} = -\sum_{i \in \mathcal{M}} \log p_{\theta_{mask}}\!\left(\mathbf{c}_i \mid \tilde{\mathbf{c}},\, \mathbf{e}_c\right),
\end{equation}
where $\mathcal{M}$ denotes the set of masked positions and $\tilde{\mathbf{c}}$ the masked token sequence. At inference, the remasking strategy extends this semantic awareness by combining semantic and prediction confidence scores rather than relying on confidence alone:
\begin{equation}
    \mathcal{R}_i = \alpha \cdot \mathcal{R}^{\text{sem}}_i + \beta \cdot \mathcal{R}^{\text{conf}}_i,
\end{equation}
where $\mathcal{R}^{\text{sem}}_i = \bar{q}_i \cdot (1-t)$ is the semantic score and $\mathcal{R}^{\text{conf}}_i$ is the token prediction confidence, ensuring that tokens with both low semantic relevance and low prediction confidence are remasked first.

\subsubsection{Style Residual Transformer}
\label{method:stage2-2}
To generate style tokens $\mathbf{s}^{1:N}$ for the remaining 
$N$ residual layers, we train a Style Residual Transformer 
(SRT) conditioned on the predicted content tokens and a 
style reference prompt $\mathbf{m}_r$. The reference style tokens 
$\mathbf{s}^{1:N}_r$ are extracted from $\mathbf{m}_r$ via 
the trained RVQ-VAE encoder. The SRT predicts one residual 
layer at a time: at layer $j$, the style tokens from 
all preceding layers $1, \ldots, j-1$ and the content tokens $\mathbf{c}$ from the first layer are embedded and summed as input, 
and the style reference tokens $\mathbf{s}^{1:N}_r$ are 
embedded and summed as the conditioning signal. During 
training, a target layer $j$ is randomly sampled from 
$\{1, \ldots, N\}$, and the SRT is optimized via:
\begin{equation}
    \mathcal{L}_{\text{res}} = \mathbb{E}_{j \sim U(1,N)}\left[\sum_{i=1}^{T'} 
    -\log p_{\theta_{res}}\!\left(\mathbf{s}^{(j)}_i \mid \mathbf{s}^{1:j-1}_i,\, \mathbf{s}^{1:N}_r,\, 
    \mathbf{c}_i,\, j\right)\right].
\end{equation}


%% file: sections/exp.tex
\subsection{Dataset}\label{subsec:dataset}

We train and evaluate our model using the BEAT2 \citep{liu2024emage} dataset, which comprises 60 hours of high-quality SMPL-based gesture data collected from 25 speakers (12 female and 13 male). For consistency, we adopt the same train-validation-test split protocol as prior work \citep{liu2024emage}. To evaluate the model's ability to generalize across speakers, we use data from 20 speakers for training and validation, and report results of seen-speaker evaluation on held-out test sequences from these 20 speakers, and zero-shot unseen-speaker evaluation on test sequences from the remaining 5 speakers.

\subsection{Evaluation Metrics}\label{exp:metrics}

We evaluate both reconstruction and generation quality. For VAE reconstruction, we report \textbf{Joints Rotation Mean Square Error (JRMSE)}~\citep{yinpyramotion} across four body regions (face, upper body, hands, lower body) and their weighted aggregate, alongside whole-body \textbf{Mean Squared Error (MSE)} and \textbf{L1 Vertex Difference (LVD)}~\citep{codetalker} for mesh-level accuracy. Distributional fidelity is measured by \textbf{Fr\'{e}chet Gesture Distance (FGD)}~\citep{yoon2020speech} with two encoders: our RVQ-VAE (\textbf{FGD}) and VAESKConv (\textbf{FGD$_{\text{sk}}$})~\citep{liu2024emage}. \textbf{Diversity}~\citep{audio2gestures} evaluates motion variation and \textbf{Normalized Beat Constancy (NBC)}~\citep{li2021ai} measures rhythmic fidelity. For generation, we report FGD, Diversity, \textbf{Beat Constancy (BC)}~\citep{li2021ai}, MSE, LVD, and facial \textbf{FaceMSE}/\textbf{FaceLVD}~\citep{codetalker}. All results are mean $\pm$ std over five runs, with statistical significance assessed via the Wilcoxon signed-rank test~\citep{wilcoxon1992individual}.



%% file: sections/results.tex
\subsection{Quantitative Results}
\label{results:QR}

\paragraph{Co-speech Gesture Generation Benchmark.}

We evaluate PersonaGest against state-of-the-art co-speech gesture generation methods, including EMAGE~\citep{liu2024emage}, MambaTalk~\citep{xu2024mambatalk}, EchoMask~\citep{zhang2025echomask}, SemTalk~\citep{zhang2025semtalk}, PyraMotion~\citep{yinpyramotion}, and GestureLSM~\citep{liu2025gesturelsm}. As shown in Table~\ref{tab:unconditional}, under the zero-shot unseen speaker setting, PersonaGest achieves the best FGD, FGD$_\text{sk}$, BC, and LVD, indicating a motion distribution closer to the ground truth, stronger speech-motion synchronization, and smoother generation. Most baselines suffer substantial FGD$_\text{sk}$ degradation on unseen speakers, whereas PersonaGest remains stable, demonstrating stronger cross-speaker generalization. Methods with strong BC or Diversity scores, such as MambaTalk and EchoMask, still show larger deviations from the ground-truth motion distribution, suggesting a trade-off between perceptual diversity and motion fidelity. Seen speaker results are provided in the supplementary materials.

{
\setlength{\heavyrulewidth}{1.5pt}
\setlength{\lightrulewidth}{0.45pt}
\begin{table*}[t]
\vspace{-2mm}
\centering
\caption{Quantitative comparison with state-of-the-art co-speech gesture generation models under zero-shot unseen speaker settings. For clarity, we report FGD$\times 10^{-5}$, FGD$_\text{sk}$$\times 10^{-1}$, MSE$\times 10^{-6}$, and LVD$\times 10^{-2}$, FaceMSE$\times 10^{-8}$, and FaceLVD$\times 10^{-5}$. \textbf{Bold}: best; \underline{underline}: second best. $^\dagger$GestureLSM does not generate facial parameters; `---' indicates metric not available.}
\label{tab:unconditional}
\resizebox{\textwidth}{!}{%
\begin{tabular}{l cccc cccc}
\toprule
\textbf{Model}
  & \textbf{FGD} $\downarrow$
  & \textbf{FGD$_\text{sk}$} $\downarrow$
  & \textbf{BC} $\uparrow$
  & \textbf{Diversity} $\uparrow$
  & \textbf{MSE} $\downarrow$
  & \textbf{LVD}  $\downarrow$
  & \textbf{FaceMSE} $\downarrow$
  & \textbf{FaceLVD} $\downarrow$ \\
\midrule
EMAGE~(CVPR'24)
  & 3.667 & 4.080 & \underline{0.812}\scriptsize{±0.024} & \underline{12.116}
  & 0.980\scriptsize{±0.280} & 6.010\scriptsize{±0.930} & 6.910\scriptsize{±1.200} & 8.120\scriptsize{±0.710} \\
MambaTalk~(NeurIPS'24)
  & 3.706 & \underline{3.752} & 0.807\scriptsize{±0.022} & 10.743
  & 0.810\scriptsize{±0.170} & 5.630\scriptsize{±0.750} & 7.240\scriptsize{±1.900} & 8.160\scriptsize{±1.200} \\
EchoMask~(MM'25)
  & 3.172 & 5.268 & 0.800\scriptsize{±0.012} & \textbf{13.881}
  & 1.980\scriptsize{±0.780} & 8.920\scriptsize{±1.510} & 7.360\scriptsize{±1.300} & 8.570\scriptsize{±0.850} \\
SemTalk~(ICCV'25)
  & 3.578 & 4.296 & 0.807\scriptsize{±0.024} & 11.788
  & 1.010\scriptsize{±0.260} & 6.390\scriptsize{±1.050} & 7.790\scriptsize{±1.600} & 8.750\scriptsize{±1.000} \\
PyraMotion~(NeurIPS'25)
  & \underline{2.411} & 3.761 & 0.678\scriptsize{±0.151} & 8.637
  & 0.840\scriptsize{±0.320} & 5.350\scriptsize{±1.280} & \textbf{4.630}\scriptsize{±2.100} & \underline{6.630}\scriptsize{±1.400} \\
GestureLSM$^\dagger$~(ICCV'25)
  & 2.949 & 3.928 & 0.725\scriptsize{±0.053} & 7.600
  & \textbf{0.590}\scriptsize{±0.100} & \underline{4.900}\scriptsize{±0.420} & --- & --- \\
\textbf{PersonaGest~(Ours)}
  & \textbf{2.311} & \textbf{2.660} & \textbf{0.826}\scriptsize{±0.040} & 11.970
  & \underline{0.780}\scriptsize{±0.220} & \textbf{4.630}\scriptsize{±0.840} & \underline{5.300}\scriptsize{±1.300} & \textbf{6.100}\scriptsize{±1.000} \\
\midrule
\textit{p-value}
  & <0.0001 & <0.0001 & <0.0001 & <0.0001
  & <0.0001 & <0.0001 & <0.0001 & <0.0001 \\
\bottomrule
\end{tabular}%
}
\vspace{-1mm}
\end{table*}
}

\paragraph{Style-Conditioned Co-speech Gesture Generation.} 
We compare PersonaGest with two style-conditioned co-speech generation models  SynTalker~\citep{chen2024enabling} and ZeroEGGS~\citep{ghorbani2023zeroeggs}, that support motion style prompts. As shown in Table~\ref{tab:conditional}, under the zero-shot unseen speaker setting, FGD and FGD$_\text{sk}$ are computed against the style reference motions rather than the ground-truth test set, directly measuring prompt-style consistency. PersonaGest achieves the best FGD, FGD$_\text{sk}$, and BC by a clear margin, while maintaining the second-best Diversity, indicating stronger style fidelity, speech-beat alignment, and expressiveness. In contrast, ZeroEGGS suffers from severe Diversity collapse, and SynTalker shows substantial degradation in distribution quality. Overall, PersonaGest better balances style fidelity, motion naturalness, and cross-speaker generalization. Seen speaker results are provided in the supplementary materials.

{
\setlength{\heavyrulewidth}{1.5pt}
\setlength{\lightrulewidth}{0.45pt}
\begin{table*}[t]
\centering
\caption{Quantitative comparison with style-conditioned co-speech gesture generation models zero-shot unseen speaker settings. For clarity, we report FGD$\times 10^{-1}$, FGD$_\text{sk}$$\times 10^{-1}$, MSE$\times 10^{-5}$, and LVD$\times 10^{-2}$, FaceMSE$\times 10^{-8}$, and FaceLVD$\times 10^{-5}$. \textbf{Bold}: best; \underline{underline}: second best. $^\dagger$SynTalker does not generate facial parameters; `---' indicates metric not available.}
\label{tab:conditional}
\resizebox{\textwidth}{!}{%
\begin{tabular}{l cccc cc cc}
\toprule
\textbf{Model}
  & \textbf{FGD} $\downarrow$
  & \textbf{FGD$_\text{sk}$} $\downarrow$
  & \textbf{BC} $\uparrow$
  & \textbf{Diversity} $\uparrow$
  & \textbf{MSE} $\downarrow$
  & \textbf{LVD} $\downarrow$
  & \textbf{FaceMSE} $\downarrow$
  & \textbf{FaceLVD} $\downarrow$ \\
\midrule
SynTalker$^\dagger$
  & 3.268 & \underline{3.242} & 0.687\scriptsize{±0.086} & \textbf{10.233}
  & 8.000\scriptsize{±1.900} & 5.740\scriptsize{±0.850} & --- & --- \\
ZeroEGGS
  & \underline{2.875} & 3.779 & \underline{0.717}\scriptsize{±0.014} & 3.274
  & \textbf{4.900}\scriptsize{±2.200} & \textbf{4.200}\scriptsize{±1.040} & \underline{8.460}\scriptsize{±2.400} & \underline{9.640}\scriptsize{±1.100} \\
\textbf{PersonaGest~(Ours)}
  & \textbf{2.617} & \textbf{2.726} & \textbf{0.815}\scriptsize{±0.034} & \underline{8.985}
  & \underline{7.900}\scriptsize{±2.100} & \underline{5.270}\scriptsize{±0.740} & \textbf{5.450}\scriptsize{±1.500} & \textbf{7.340}\scriptsize{±0.950} \\
\midrule
\textit{p-value}
  & <0.0001 & <0.0001 & <0.0001 & <0.0001
  & >0.1 & >0.1 & <0.0001 & <0.0001 \\
\bottomrule
\end{tabular}%
}
\vspace{-3mm}
\end{table*}
}

\subsection{Perceptual Study}\label{results:PS}
\begin{figure*}[t]
  \centering
  \begin{subfigure}[b]{0.49\textwidth}
    \includegraphics[width=\textwidth]{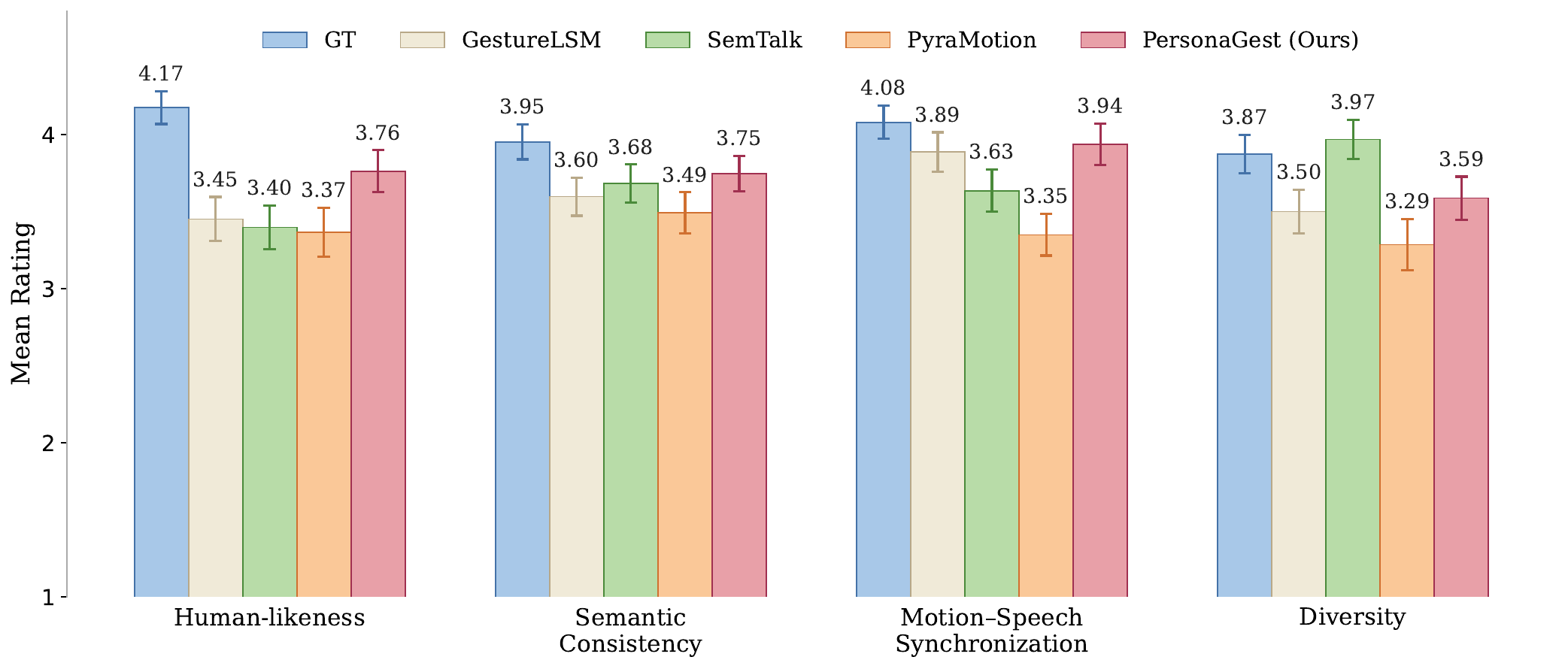}
    \caption{Co-speech gesture generation benchmark.}
    \label{fig:perceptual_part1}
  \end{subfigure}
  \hfill
  \begin{subfigure}[b]{0.49\textwidth}
    \includegraphics[width=\textwidth]{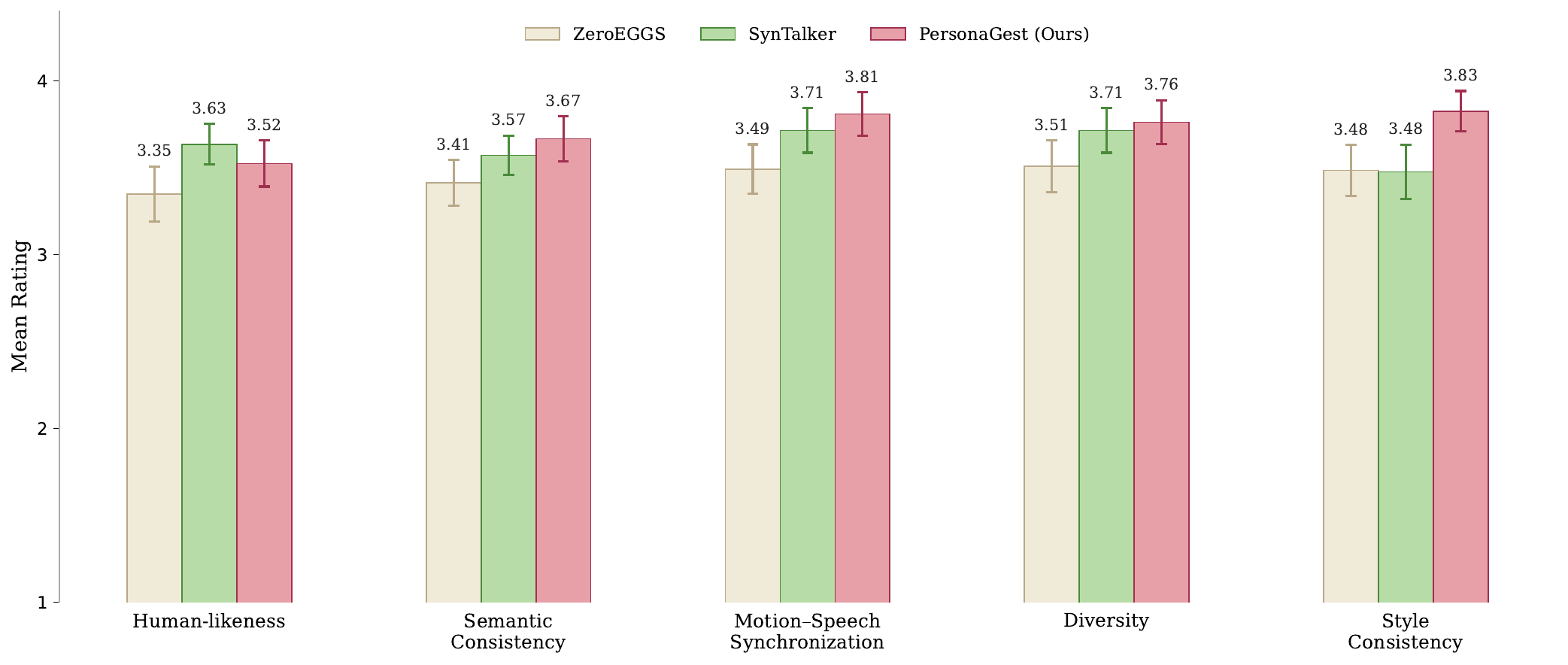}
    \caption{Style-conditioned co-speech gesture generation.}
    \label{fig:perceptual_part2}
  \end{subfigure}
  \caption{Perceptual study results on a 5-point Likert-like scale (higher is better). \textbf{Human-likeness}: resemblance to natural human motion. \textbf{Semantic consistency}: alignment between gestures and speech content. \textbf{Motion--speech synchronization}: temporal alignment with speech rhythm. \textbf{Diversity}: variety and expressiveness of generated motions. \textbf{Style consistency} ((b) only): preservation of the reference motion style.}
  \label{fig:user-study}
\end{figure*}


\paragraph{User Study.}
We conduct a two-part perceptual study with 21 participants. 
Part~I evaluates five test samples from all comparison methods in terms of \textit{human-likeness}, \textit{semantic consistency}, \textit{motion-speech synchronization}, and \textit{diversity}. Part~II focuses on style-conditioned methods, using the same criteria and an 
additional \textit{style consistency} rating for reference-generation pairs. As shown in Figure~\ref{fig:user-study}, PersonaGest achieves the best overall performance among generative methods, approaching GT-level scores on most 
Part~I metrics. While SemTalk obtains the highest diversity score, it also underperforms on the remaining metrics and even exceeds GT diversity, suggesting overly exaggerated rather than meaningfully expressive motion. In Part~II, 
PersonaGest consistently outperforms SynTalker and ZeroEGGS, especially in style consistency.

\begin{figure}[t!]
\centering
\includegraphics[width=1.0\textwidth]{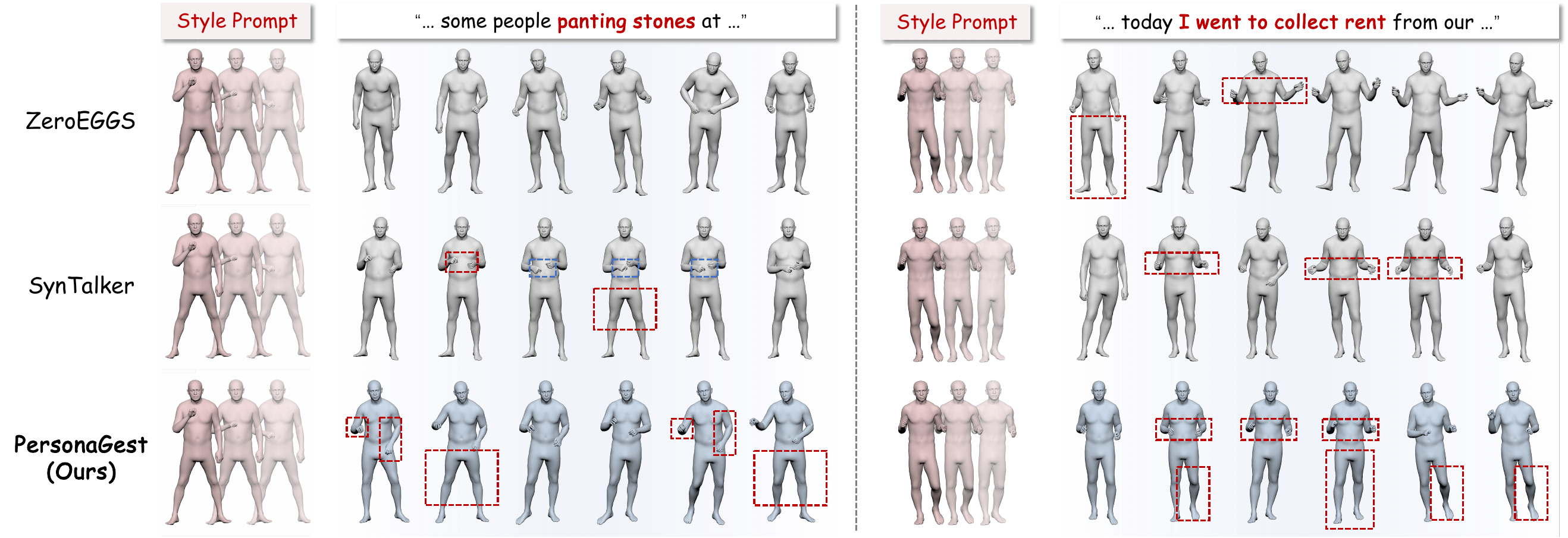}
\caption{Qualitative comparison with ZeroEGGS and SynTalker on style-conditioned gesture generation. All methods share the same style motion prompt (pink, left) and speech audio.}
\label{fig:vis-cond}
\vspace{-3mm}
\end{figure}

\paragraph{Visualization.} 
Figure~\ref{fig:vis-cond} shows qualitative comparisons across two examples. ZeroEGGS generates diverse motions but fails to preserve the style characteristics of the reference prompt, producing gestures largely independent of the given style. SynTalker captures the general trend of the reference style, yet the generated motions lack precision in reproducing the stylistic details of the prompt, and style consistency degrades over time. In contrast, PersonaGest more faithfully reproduces the stylistic features of the reference, maintaining both accurate motion amplitude and consistent style throughout the sequence, as highlighted by the red dashed boxes.

\subsection{Ablation Study}\label{results:AS}

\paragraph{Comparison of VQ-VAE Motion Representations.}
We compare PersonaGest’s RVQ-VAE tokenizer with representative VQ-based motion representations, including VQ-VAE~\citep{liu2024emage}, APVQ-VAE~\citep{yinpyramotion}, RVQ-VAE (\textbf{S})~\citep{zhang2025semtalk}, and RVQ-VAE (\textbf{B})~\citep{liu2025gesturelsm}. As shown in Table~\ref{tab:vq}, under the zero-shot unseen speaker setting, PersonaGest achieves top-2 performance on nearly all metrics, indicating strong cross-speaker generalization. In contrast, RVQ-VAE (S) excels only in Face reconstruction and FGD, while RVQ-VAE (B) shows less consistent performance. Overall, PersonaGest offers a more balanced trade-off among reconstruction fidelity, motion naturalness, and generalization. Seen speaker results are provided in the supplementary materials.

{
\setlength{\heavyrulewidth}{1.5pt}
\setlength{\lightrulewidth}{0.45pt}
\begin{table*}[t]
\vspace{-2mm}
\centering
\caption{Quantitative comparison of VQ-based motion representation models under zero-shot unseen speaker settings. For clarity, we report Face$\times 10^{-3}$, Upper$\times 10^{-2}$, Hands$\times 10^{-2}$, Lower$\times 10^{-2}$, JRMSE$\times 10^{-2}$, MSE$\times 10^{-5}$, LVD$\times 10^{-2}$, FGD$\times 10^{-3}$ and FGD$_\text{sk}$$\times 10^{-1}$. \textbf{Bold}: best; \underline{underline}: second best.}
\label{tab:vq}
\resizebox{\textwidth}{!}{%
\begin{tabular}{l ccccc cc cccc}
\toprule
\textbf{Model}
  & \textbf{Face $\downarrow$}
  & \textbf{Upper$\downarrow$}
  & \textbf{Hands$\downarrow$}
  & \textbf{Lower$\downarrow$}
  & \textbf{JRMSE$\downarrow$}
  & \textbf{MSE $\downarrow$}
  & \textbf{LVD $\downarrow$}
  & \textbf{FGD $\downarrow$}
  & \textbf{FGD$_\text{sk}$ $\downarrow$}
  & \textbf{NBC $\downarrow$}
  & \textbf{Diversity $\uparrow$}\\

\midrule
VQ-VAE
  & 1.065\scriptsize{±0.241} & 1.524\scriptsize{±0.593} & 2.520\scriptsize{±1.343} & 1.687\scriptsize{±1.144} & 1.637\scriptsize{±0.735}
  & 3.900\scriptsize{±1.150} & 3.850\scriptsize{±0.665}
  & 6.898 & 2.530 & 1.971\scriptsize{±0.366} & 5.138 \\
APVQ-VAE
  & 0.777\scriptsize{±0.283} & 0.800\scriptsize{±0.392} & 1.437\scriptsize{±0.722} & 1.111\scriptsize{±1.072} & 0.933\scriptsize{±0.487}
  & 3.110\scriptsize{±1.560} & 3.520\scriptsize{±1.010}
  & 3.179 & 1.165 & 2.122\scriptsize{±0.269} & 6.289 \\
RVQ-VAE (S)
  & \textbf{0.580}\scriptsize{±0.035} & \textbf{0.276}\scriptsize{±0.179} & \underline{0.599}\scriptsize{±0.313} & \underline{0.227}\scriptsize{±0.126} & \underline{0.345}\scriptsize{±0.166}
  & \underline{0.814}\scriptsize{±0.314} & \underline{1.890}\scriptsize{±0.390}
  & \textbf{0.904} & \underline{0.608} & 1.791\scriptsize{±0.320} & \textbf{6.817} \\
RVQ-VAE (B)
  & \underline{0.629}\scriptsize{±0.038} & 0.304\scriptsize{±0.229} & 0.735\scriptsize{±0.358} & 0.302\scriptsize{±0.185} & 0.421\scriptsize{±0.192}
  & 1.200\scriptsize{±0.489} & 2.310\scriptsize{±0.552}
  & 2.280 & 0.883 & \underline{1.397}\scriptsize{±0.384} & \underline{6.507} \\
\textbf{PersonaGest}
  & 0.719\scriptsize{±0.219} & \underline{0.298}\scriptsize{±0.196} & \textbf{0.501}\scriptsize{±0.269} & \textbf{0.226}\scriptsize{±0.144} & \textbf{0.326}\scriptsize{±0.156}
  & \textbf{0.809}\scriptsize{±0.360} & \textbf{1.880}\scriptsize{±0.460}
  & \underline{0.956} & \textbf{0.517} & \textbf{1.389}\scriptsize{±0.324} & 6.389 \\
  \midrule
\textit{p-value}
  & <0.0001 & <0.0001 & <0.0001 & <0.0001 & <0.0001
  & <0.0001 & <0.0001
  & <0.0001 & <0.0001 & <0.0001 & <0.0001 \\
\bottomrule
\end{tabular}%
}
\end{table*}
\vspace{-3mm}
}

{
\setlength{\heavyrulewidth}{1.5pt}
\setlength{\lightrulewidth}{0.45pt}
\begin{table*}[t!]
\vspace{-2mm}
\centering
\caption{Ablation study on key components of PersonaGest. The upper block ablates RVQ-VAE components, where NBC is reported for rhythmic fidelity and lower values are better; the lower block ablates the Stage 2 generation component, where BC is reported for speech-motion synchronization and higher values are better. The left and right parts report results on seen and zero-shot unseen speakers, respectively.}
\label{tab:ablation_components}
\resizebox{\textwidth}{!}{%
\begin{tabular}{l cccc | cccc}
\toprule
\textbf{Variant}
  & \textbf{FGD} $\downarrow$
  & \textbf{FGD$_\text{sk}$} $\downarrow$
  & \textbf{NBC}$\downarrow$ / \textbf{BC}$\uparrow$
  & \textbf{Diversity} $\uparrow$
  & \textbf{FGD} $\downarrow$
  & \textbf{FGD$_\text{sk}$} $\downarrow$
  & \textbf{NBC}$\downarrow$ / \textbf{BC}$\uparrow$
  & \textbf{Diversity} $\uparrow$ \\
\midrule
-w/o $\mathbf{Q}_{\rm learn}$    & 1.872 & 0.714 & 1.078\scriptsize{±0.477} & 6.863 & 1.142 & 6.622 & 1.461\scriptsize{±0.307} & 5.489 \\
-w/o MSAF  & 1.945 & 0.820 & 1.057\scriptsize{±0.468} & 5.842 & 1.131 & 5.982 & 1.392\scriptsize{±0.361} & 5.445 \\
-w/o SMoC           & 2.082 & 0.944 & 1.042\scriptsize{±0.472} & 6.854 & 1.193 & 7.630 & 1.463\scriptsize{±0.379} & 4.460 \\
-w/o CPA          & 1.965 & 0.634 & 1.068\scriptsize{±0.473} & 6.367 & 1.059 & 7.323 & 1.470\scriptsize{±0.286} & 5.522 \\
\textbf{RVQ-VAE (Ours)} & \textbf{1.787} & \textbf{0.483} & \textbf{1.024}\scriptsize{±0.484} & \textbf{8.302} & \textbf{0.956} & \textbf{5.169} & \textbf{1.389}\scriptsize{±0.324} & \textbf{6.389} \\
\midrule
-w/o SAM                      & 0.277 & 1.430 & 0.607\scriptsize{±0.086} & 10.818 & 2.668 & 2.710 & 0.776\scriptsize{±0.036} & 11.332 \\
\textbf{PersonaGest (Ours)}     & \textbf{0.248} & \textbf{1.414} & \textbf{0.859}\scriptsize{±0.091} & \textbf{11.053} & \textbf{2.311} & \textbf{2.660} & \textbf{0.826}\scriptsize{±0.040} & \textbf{11.970} \\
\bottomrule
\end{tabular}%
}
\vspace{-3mm}
\end{table*}
}

\paragraph{Key Components.} 
We ablate key components of PersonaGest across both stages. In Stage~1, removing the Semantic-Aware Motion Codebook (SMoC) causes the largest FGD degradation in both seen and zero-shot settings, highlighting the importance of semantic routing for motion-aware discrete representations. Removing the learnable global token $\mathbf{Q}_{\rm learn}$ or the multi-scale audio fusion block (MSAF) also degrades performance, especially diversity, confirming the role of audio-motion synchronization and multi-scale audio cues in expressive tokenization. Removing the cross-part attention block (CPA) further harms FGD$_\text{sk}$ and diversity, indicating the need for cross-part interaction in coherent full-body motion. In Stage~2, removing Semantic-Aware Masking (SAM) notably reduces zero-shot BC and diversity, showing that semantic guidance improves generalization to unseen speakers.

\begin{wraptable}{r}{0.56\columnwidth}
\vspace{-1pt}
\centering

\captionsetup{
    font=small,
    labelfont=normalfont,
    justification=raggedright,
    singlelinecheck=false,
    skip=3pt
}

\captionof{table}{Speaker identification accuracy (\%) across ablation variants.}
\label{tab:probe_ablation}

\renewcommand{\arraystretch}{1.03}
\scriptsize
\setlength{\tabcolsep}{1.6pt}

\vspace{2pt}

\begin{adjustbox}{max width=\linewidth}
\begin{tabular}{@{}lcccc@{\hspace{2pt}}|@{\hspace{2pt}}cccc@{}}
\toprule[1.1pt]
& \multicolumn{4}{c}{\textbf{Content$\to$Speaker}} 
& \multicolumn{4}{c}{\textbf{Style$\to$Speaker}} \\
\textbf{Variant} 
& Upper & Hands & Lower & All 
& Upper & Hands & Lower & All \\
\midrule
-w/o $\mathcal{L}_{\rm cl}$          
& 11.4 & 10.3 & 9.1 & 9.7 
& 50.6 & 56.6 & 52.8 & 49.5 \\
-w/o $\mathcal{L}_{\rm cl}$+$\mathcal{L}_{\rm phone}$  
& 29.0 & 32.4 & 24.2 & 17.6 
& 60.4 & 46.6 & 68.4 & 54.2 \\
\textbf{RVQ-VAE (Ours)}       
& 13.8 & 17.2 & 13.9 & 12.1 
& 86.3 & 61.5 & 98.6 & 78.9 \\
\bottomrule[1.1pt]
\end{tabular}
\end{adjustbox}

\vspace{4pt}

\includegraphics[width=0.49\linewidth]{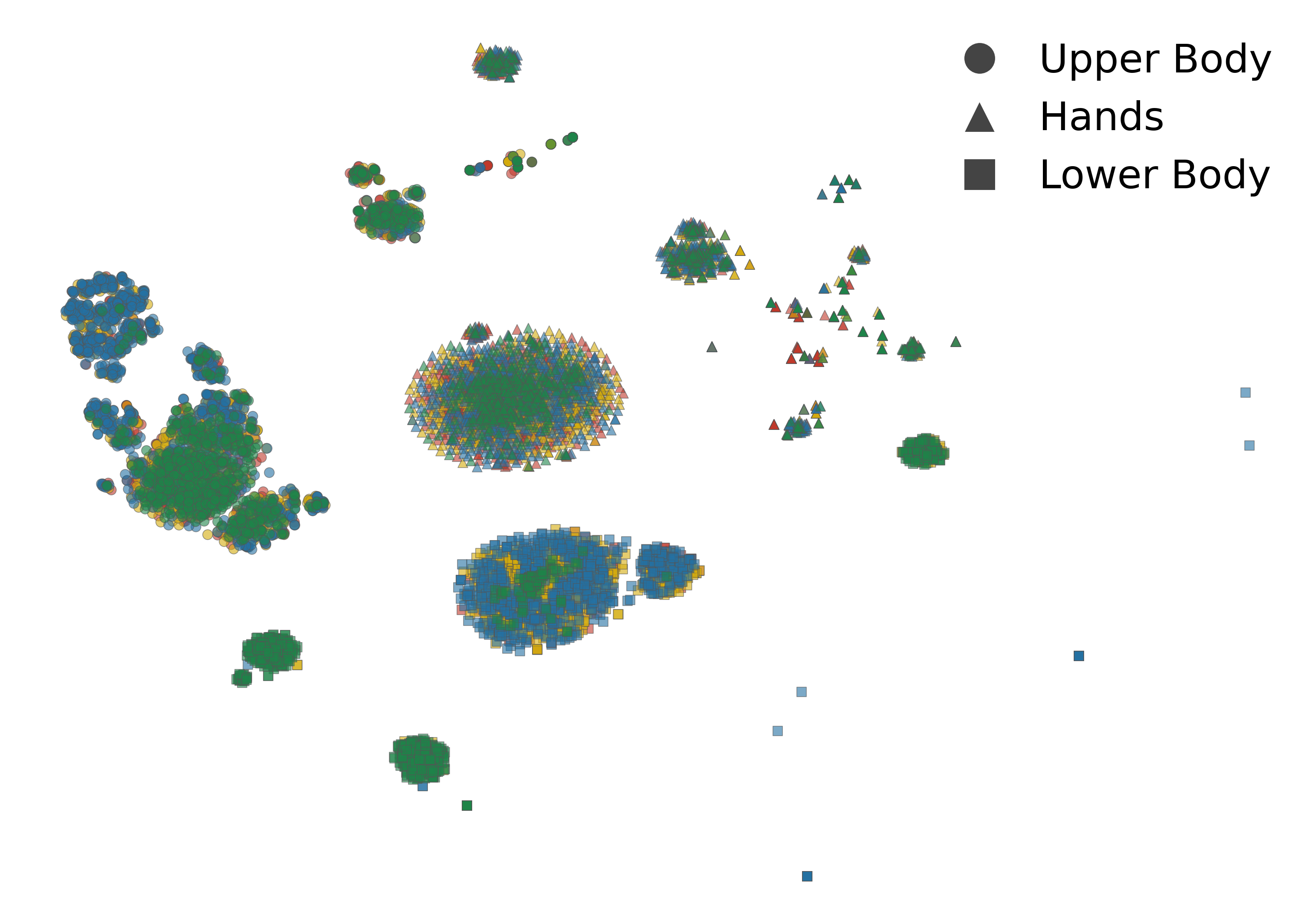}
\includegraphics[width=0.49\linewidth]{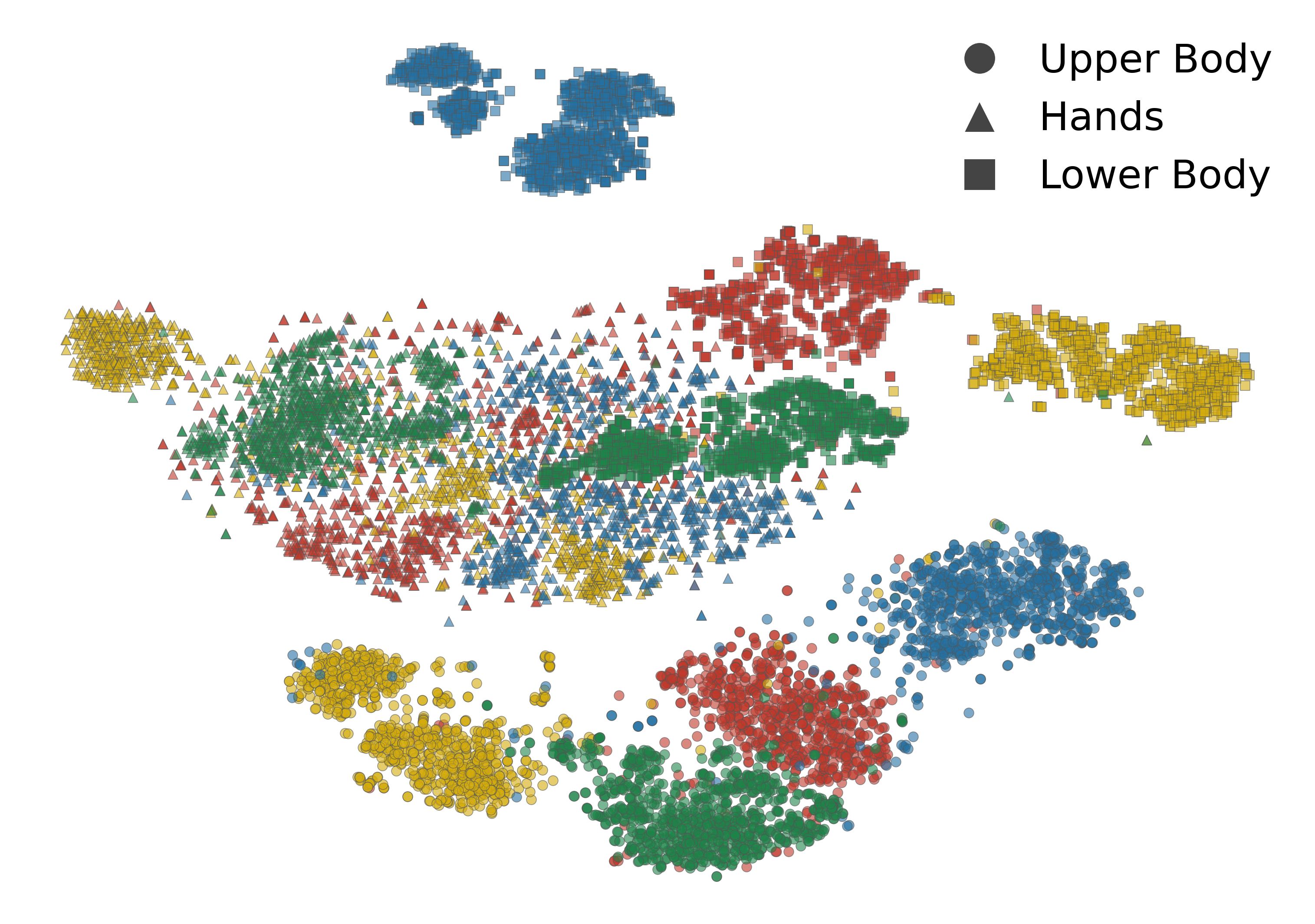}

\captionof{figure}{T-SNE visualization of content (left) and style (right) embeddings. 
Colors denote different speakers.}
\label{fig:disentangle}

\vspace{-6pt}
\end{wraptable}

\paragraph{Disentanglement of Content and Style Representations.}

We ablate two key training components to assess their roles in style learning: the style contrastive loss $\mathcal{L}_{\rm cl}$, and the combination of phoneme supervision $\mathcal{L}_{\rm phone}$ and $\mathcal{L}_{\rm cl}$. As shown in Table~\ref{tab:probe_ablation}, we first train a linear classifier to classify speaker identity from frozen content and style latents separately. Our full model achieves content accuracy near chance across all body parts while style latents yield high speaker classification accuracy, confirming effective disentanglement. Removing $\mathcal{L}_{\rm cl}$ substantially reduces style clustering, while removing $\mathcal{L}_{\rm phone}$ causes speaker identity to leak into the content space, validating the distinct role of each objective. Additionally, as shown in Figure~\ref{fig:disentangle}, we visualize the content and style embeddings via t-SNE. Content representations are interleaved across speakers, indicating speaker-agnostic motion encoding, whereas style representations form well-separated speaker clusters, demonstrating effective capture of speaker-specific motion characteristics.

%% file: sections/concl.tex
\vspace{-3mm}
We present PersonaGest, a two-stage framework for style-conditioned co-speech 3D gesture generation. The first stage introduces an RVQ-VAE-based motion tokenizer that learns compact discrete representations of full-body gestures, outperforming existing VQ-based alternatives in both reconstruction fidelity and motion naturalness. The second stage employs a style conditioning mechanism that takes a reference motion clip as a prompt, enabling controllable gesture generation that preserves the stylistic characteristics of the reference speaker while maintaining speech alignment. Extensive experiments on BEAT2 demonstrate state-of-the-art performance across both seen and zero-shot speaker settings, under unconditional and style-conditioned evaluations. Perceptual study results further confirm that PersonaGest generates gestures that are more natural, semantically consistent, and style-faithful than existing approaches.

%% file: sections/appendix.tex

\startcontents[appendix]

\begin{center}
    {\Large\bfseries
    PersonaGest: Personalized Co-Speech Gesture Generation with\\
    Semantic-Guided Hierarchical Motion Representation\par}
    \vspace{0.8em}
    {\Large\bfseries Appendices\par}
\end{center}

\vspace{1.2em}

\printcontents[appendix]{}{1}{
    \setcounter{tocdepth}{2}
}

\clearpage


\section{Related Work on VQ-based Motion Representation}
\label{ap:Related Work}
\input{sections/appendix/related_work}

\section{Implementation Details}
\label{ap:Implementation Details}
\input{sections/appendix/implement_details}

\section{Efficiency Analysis}
\label{ap:Efficiency Analysis}
\input{sections/appendix/efficiency}

\section{Objective Evaluation Metrics}
\label{ap:Objective Evaluation Metrics}
\input{sections/appendix/metrics}

\section{Subjective Listening Test}
\label{ap:Subjective Listening Test}
\input{sections/appendix/user_study}

\section{Supplementary Experimental Results}
\label{ap:Experimental Results}
\input{sections/appendix/sup_results}

\section{Supplementary Rendering Results}
\label{ap:RS}
\input{sections/appendix/sup_vis}

\section{Limitations and Future Work}
\label{ap:limitation}
\input{sections/appendix/limit}

\section{Broader Impacts}
\label{ap:impact}
\input{sections/appendix/broader_impact}

%% file: sections/appendix/related_work.tex
Learning compact and expressive motion representations is fundamental to high-quality motion generation. Early VAE-based methods \citep{petrovich2021action, guo2022generating, liu2026holisticsemges, jian2024learn, yazdian2022gesture2vec} capture the overall motion distribution in a continuous latent space. VQ-VAE-based methods \citep{liu2022audio, yi2023generating, zhong2023attt2m} further improve this by encoding motion into discrete token sequences, enabling token-based generation with improved reconstruction fidelity. However, single-codebook quantization accumulates reconstruction errors, motivating the use of Residual VQ-VAE \citep{guo2024momask, liu2025gesturelsm}, which progressively refines quantization residuals through hierarchical codebooks. Beyond quantization depth, part-based representations have been explored to handle the heterogeneous nature of full-body motion: EMAGE \citep{liu2024emage} employs separate VQ-VAEs per body part, and PyraMotion \citep{yinpyramotion} encodes motion patterns at multiple temporal scales. VQ-Style~\citep{zargarbashi2026vq} builds on RVQ-VAE to disentangle motion content and style across the codebook hierarchy via contrastive learning and mutual information loss, enabling zero-shot style transfer through codebook swapping at inference time. However, VQ-Style is designed for general motion sequences rather than co-speech gesture generation, and does not incorporate any semantic information into the representation, leaving gesture semantics unaddressed in the learned tokens.

%% file: sections/appendix/implement_details.tex
\subsection{Model Details}
\label{ap:Model Details}
\begin{figure}
\vspace{-2mm}
\centering
\includegraphics[width=0.9\textwidth]{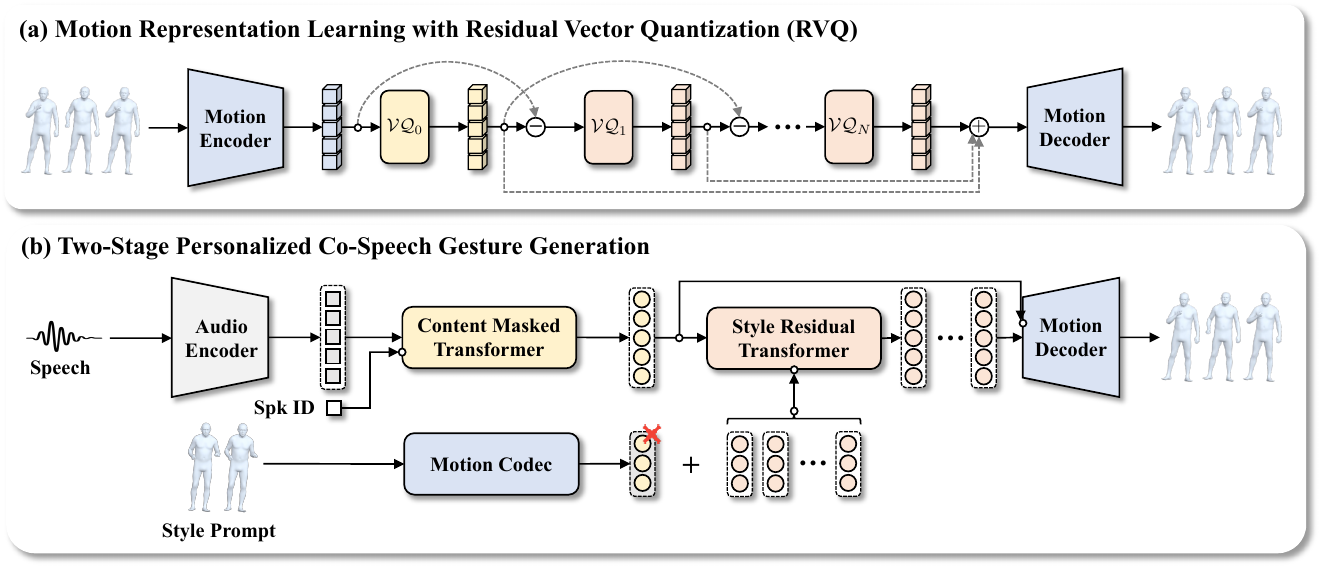}
\caption{\textbf{Overview of PersonaGest.} \textbf{Stage 1:} A semantic-guided RVQ-VAE encodes motion into disentangled content and style latent codes. \textbf{Stage 2:} A Content Masked Transformer generates content tokens conditioned on speech and speaker identity, followed by a Style Residual Transformer that generates style tokens conditioned on a reference motion prompt. \label{fig:overview}}
\vspace{-3mm}
\end{figure}

Figure~\ref{fig:overview} provides an overview of PersonaGest. We describe the architectural details of each component below.

\paragraph{RVQ-VAE.}
The encoder and decoder use a 1D convolutional architecture with temporal downsampling factor $4\times$, width $512$, depth $3$, and dilation growth rate $3$. The codebook has size $N_c = 512$ with code dimension $d = 128$, organized into $N = 8$ residual layers: $1$ content layer and $7$ style layers. The SMoC partitions the content codebook into $K = 5$ semantic regions of $102$ entries each (total $510$ slots), with temperature sharpness $\kappa = 1.0$ and EMA decay $\mu = 0.99$. The MSAF module projects audio features through a shared prosody projection (LayerNorm $\to$ Linear $\to$ GELU $\to$ Linear)~\citep{lyu2026ksdiff, li2025prosodytalker, zhao2025prosody}, fuses multi-scale prosody at $1\times$, $2\times$,  $4\times$ resolutions via an MLP before cross-attention, and injects attended features via per-part gated residuals with scalar gates initialized to $-2.0$ (hands) and $-3.0$ (others) to suppress early-training influence. The CPA module uses $1$ post-quantization Transformer layer with $4$ attention heads. The phoneme predictor $\varphi(\cdot)$, inspired by NaturalSpeech3~\citep{ju2024naturalspeech}\footnote{\url{https://github.com/lifeiteng/naturalspeech3_facodec}}, 
takes the motion content latents as the input, followed by three 
dilated residual units (dilation $=1,2,3$) with SnakeBeta activations~\citep{lee2022bigvgan} 
and weight-normalized convolutions, and a linear head over $71$ ARPABET categories.


\paragraph{Content Masked Transformer.}
The CMT consists of $8$ Transformer encoder layers with $6$ attention heads, latent dimension $384$, feedforward size $1{,}024$, and dropout $0.2$. Speech is encoded by a frozen Whisper-base encoder~\citep{akec2023robust}\footnote{\url{https://github.com/openai/whisper}} and projected to the latent dimension; speaker identity is provided as a learned embedding. CFG dropout probability is $0.2$. At inference, we run $18$ iterative decoding steps with CFG scale $4.0$ and top-$k$ filtering threshold $0.9$, with semantic remasking weights $\alpha = 0.6$,  $\beta = 0.4$, and $r = 0.2$.

\paragraph{Style Residual Transformer.}
The SRT shares the same Transformer architecture as the CMT ($8$ layers, $6$ heads, latent dimension $384$, feedforward size $1{,}024$). The style condition is derived randomly from the $25\% - 50\%$ frames of the reference motion encoded via the RVQ-VAE, with a global style token extracted through cross-attention over the style prefix. CFG dropout probability is $0.1$, and CFG scale at inference is $3.0$.

\subsection{Training Details}
\label{ap:Training Details}
We utilize the Adam optimizer with weight decay $10^{-4}$, $\beta_1{=}0.9$, $\beta_2{=}0.99$ for both training stages. The RVQ-VAE is trained for 20K iterations with learning rate $2\times10^{-4}$ and batch size $128$. For Stage~2, we freeze the RVQ-VAE and train the CMT for 20 epochs with learning rate $2\times10^{-4}$, decayed by $10\times$ at epoch~14, and the SRT for 120 epochs with learning rate $8\times10^{-4}$, decayed at epoch~80. Gradient norms are clipped to $1.0$. Motion sequences are processed at $30$\,fps using $128$-frame windows with a stride of $20$ frames. All experiments are conducted on NVIDIA A100 GPUs. For all baseline methods, we train the publicly available implementations 
on the same BEAT2 data split (20 speakers) and preprocessing pipeline to ensure fair comparison.

\subsection{Long-Sequence Inference Details}
\label{sec:inference}
Generating gestures for long speech recordings requires handling sequences beyond PersonaGest's fixed-length context. We adopt a sliding-window inference strategy that processes audio sequentially while preserving motion continuity across boundaries.

\paragraph{Window Configuration.}
The input is partitioned into windows of $W=128$ frames ($32$ content tokens at $4:1$ compression), shifted by stride $S=96$ frames with overlap $O=32$ frames (8 tokens). The $k$-th window spans $[kS,\,kS+W)$, and the total window count is:
\begin{equation}
  K = \left\lceil\frac{T - W}{S}\right\rceil + 1.
  \label{eq:nwin}
\end{equation}

\paragraph{Autoregressive Prefix Conditioning.}
The first window ($k=0$) is decoded via standard iterative masked decoding over all $32$ tokens. For $k>0$, the final $8$ content tokens of window $k-1$ serve as a locked prefix, excluded from re-masking, while the remaining $24$ tokens are generated conditioned on this context. Style tokens $\mathbf{s}^{1:N}$ are generated independently per window without prefix conditioning.

\paragraph{Overlap-Add Assembly.}
Decoded windows are assembled via overlap-add (OLA)~\citep{crochiere1980weighted} using a shifted Hann envelope:
\begin{equation}
  h(n) = \frac{1}{2}\!\left(1 - \cos\frac{2\pi(n+1)}{W+1}\right), \quad n = 0,\ldots,W-1,
  \label{eq:hann}
\end{equation}
which is strictly positive, preventing division by zero. The reconstructed frame at time $t$ is:
\begin{equation}
  \hat{\mathbf{m}}_t =
  \frac{\sum_{k:\,t \in [kS,\,kS+W)} \mathbf{m}^{(k)}_{t-kS}\,h(t-kS)}
       {\sum_{k:\,t \in [kS,\,kS+W)} h(t-kS)},
  \label{eq:ola}
\end{equation}
where $\mathbf{m}^{(k)}_{t-kS}$ is the $(t-kS)$-th frame of the $k$-th decoded window.

\paragraph{Post-Processing.}
Translation velocity is integrated via cumulative summation and high-pass filtered to suppress quantization artifacts and prevent positional drift in long sequences.

\subsection{Ablation Model Details}
\label{sec:ablation_impl}

To isolate the contribution of each architectural component in PersonaGest, we construct ablation variants spanning both stages of the framework. Each variant removes  one module and substitutes a simpler alternative, with all other components held constant.

\paragraph{A1: -w/o Multi-Scale Audio Fusion (MSAF).}
In \textbf{A1}, the MSAF module is removed. Rather than injecting multi-scale audio features into the content latent of each body part, the quantized content latents are passed directly into the Cross-Part Attention (CPA) module without any audio fusion. This variant isolates the contribution of multi-scale prosody fusion to motion-speech synchronization and content quality.

\paragraph{A2: -w/o Cross-Part Attention (CPA).}
In \textbf{A2}, the Cross-Part Attention module applied after MSAF is removed and replaced by an identity mapping. As a result, the quantized token sequence of each body part is forwarded downstream independently, without attending to the latent representations of any other part. This variant measures the benefit of explicit inter-part coordination in the quantized latent space.

\paragraph{A3: -w/o Semantic-Aware Motion Codebook (SMoC).}
In \textbf{A3}, SMoC is replaced by a standard single-codebook VQ layer. Rather than routing each token to a semantic partition of the codebook, every content lantent $\mathbf{z}_c$ is quantized via nearest-neighbour lookup over a single shared codebook $\mathcal{C} = \{\mathbf{e}_m\}_{m=1}^{N_c}$:
\begin{equation}
  \mathbf{z}_c = \mathbf{e}_{m^*}, \quad
  m^* = \arg\min_{m}\,\bigl\|\mathbf{z} - \mathbf{e}_m\bigr\|_2,
  \label{eq:vq}
\end{equation}
where codebook entries are updated via exponential moving average (EMA). This removes both the semantic partitioning of the codebook.

\paragraph{A4: -w/o Learnable Global Token ($\mathbf{Q}_{\rm learn}$).}
In \textbf{A4}, the learnable global token $\mathbf{Q}_{\rm learn}$ in the MSAF module is removed. For body parts $p \in \{\text{upper, lower, face}\}$, the original MSAF formulation adds $\mathbf{Q}_{\rm learn}$ to the part's content latent before forming the query for multi-head cross-attention (MHCA) with audio features. In this variant, the query is computed directly from the content latent without the global token $\mathbf{Q}^p = \mathrm{LN}(\mathbf{z}_c^p)$, 
which is then used as the query in the MHCA with audio keys and values. This variant isolates the contribution of the learnable global token in aggregating cross-part context prior to audio fusion.

\paragraph{A5: -w/o Semantic-Aware Masking and Remasking (SAM).}
In \textbf{A5}, both components of SAM in Stage~2 are disabled. During training, the semantic masking priority $\mathcal{P}^{\rm sem}_i$ is removed and tokens are masked in uniformly random order following the standard cosine schedule~\citep{chang2022maskgit}. During generation, the semantic logit bias is discarded and the remasking order is determined solely by per-token prediction confidence $\mathcal{R}^{\rm conf}_i$, with no semantic guidance on which codebook partition the generated tokens are drawn from. This variant isolates the contribution of semantic awareness in both the training masking strategy and the iterative decoding procedure.

\paragraph{A6: -w/o Style Contrastive Loss and Phoneme Supervision ($\mathcal{L}_{\rm cl}$ + $\mathcal{L}_{\rm phone}$).}
In \textbf{A6}, both disentanglement-related auxiliary losses are jointly removed. The style contrastive loss $\mathcal{L}_{\rm cl}$, which enforces speaker-discriminative style representations via an InfoNCE objective, and the phoneme prediction loss $\mathcal{L}_{\rm phone}$, which regularizes the content encoder to retain phoneme-level speech information, are both disabled. The model is therefore trained without any explicit supervision on content-style factorization beyond the reconstruction objective, providing a lower bound on disentanglement performance.

\paragraph{A7: -w/o Style Contrastive Loss ($\mathcal{L}_{\rm cl}$).}
In \textbf{A7}, only $\mathcal{L}_{\rm cl}$ is removed while $\mathcal{L}_{\rm phone}$ is retained. This variant isolates the contribution of speaker-contrastive supervision from that of phoneme-level regularization, allowing us to attribute performance differences between A6 and A7 specifically to the role of the contrastive objective in shaping the style latent space.

%% file: sections/appendix/efficiency.tex

We evaluate the computational efficiency of our pipeline by measuring the runtime of each module on a single NVIDIA A100 GPU. Times are reported in seconds per second of generated motion (s/s) and averaged over multiple test sequences. As shown in Table~\ref{tab:efficiency}, the generative transformers dominate the cost: CMT accounts for approximately 60\% and SRT a further 21\% of total runtime. The RVQVAE encoders and decoders each contribute less than 0.001 s/s, confirming that discrete tokenization adds minimal overhead. The overall pipeline runs at $0.038 \pm 0.000$ s/s. We also compare against two style-conditioned baselines. SynTalker runs at 0.770 s/s due to iterative diffusion sampling, while ZeroEGGS runs at 0.014 s/s owing to its lightweight RNN decoder without discrete tokenization.

\begin{table}[h]
\vspace{-2mm}
\centering
\caption{Runtime per second of generated motion (s/s) on a single NVIDIA A100 GPU, comparing PersonaGest against style-conditioned co-speech gesture generation baselines.}
\label{tab:efficiency}
\vspace{4pt}
\small
\begin{tabular}{lc}
\toprule
\textbf{Module} & \textbf{Run Time (s/s)} \\
\midrule
\rowcolor{gray!15} SynTalker & $0.770 \pm 0.003$ \\
\rowcolor{gray!15} ZeroEGGS & $0.014 \pm 0.000$ \\
\midrule
\multicolumn{2}{l}{\textbf{Our Method}} \\
\midrule
Audio Encoder (Whisper)  & $0.00179 \pm 0.00005$ \\
\multicolumn{2}{l}{RVQVAE Encoder} \\[1pt]
\quad -- Upper              & $0.00027 \pm 0.000005$ \\
\quad -- Hands              & $0.00023 \pm 0.000001$ \\
\quad -- Lower              & $0.00023 \pm 0.000002$ \\
\quad -- Face               & $0.00022 \pm 0.000002$ \\
\multicolumn{2}{l}{RVQVAE Decoder} \\[1pt]
\quad -- Upper              & $0.00029 \pm 0.000002$ \\
\quad -- Hands              & $0.00022 \pm 0.000001$ \\
\quad -- Lower              & $0.00023 \pm 0.000001$ \\
\quad -- Face               & $0.00022 \pm 0.000001$ \\
\multicolumn{2}{l}{Generative Transformers} \\[1pt]
\quad -- CMT                & $0.02643 \pm 0.00001$ \\
\quad -- SRT                & $0.00932 \pm 0.000006$ \\
\rowcolor{gray!15} \textbf{Total Time} & $\mathbf{0.038 \pm 0.000}$ \\
\bottomrule
\end{tabular}
\vspace{-3mm}
\end{table}

%% file: sections/appendix/metrics.tex
We evaluate both reconstruction and generation quality using the following metrics.

\subsection{Joint Rotation Mean Squared Error (JRMSE)}
\label{ap:JRMSE}
For each body region $p$, JRMSE measures the mean squared error between reconstructed and ground-truth rotation features over all $n_p = T \times D_p$ values in the sequence:
\begin{equation}
  \text{JRMSE}_p = \frac{1}{n_p}\sum_{i=1}^{n_p}(r_i - \hat{r}_i)^2,
  \label{eq:jrmse}
\end{equation}
where $r_i$ and $\hat{r}_i$ are the ground-truth and reconstructed rotation values. To obtain a single holistic score, we report a dimension-weighted aggregate:
\begin{equation}
  \text{wJRMSE} = \sum_{p \in \mathcal{B}} \frac{D_p}{\sum_{p' \in \mathcal{B}} D_{p'}} \cdot \text{JRMSE}_p,
  \label{eq:wjrmse}
\end{equation}
where $\mathcal{B} = \{\text{face, upper, hands, lower}\}$ and $D_p$ is the rotation feature dimensionality of part $p$ ($D_{\text{face}}{=}100$, $D_{\text{upper}}{=}78$, $D_{\text{hands}}{=}180$, $D_{\text{lower}}{=}57$).

\subsection{Mesh Vertex Error (MSE and LVD)}
\label{ap:MSE and LVD}
To assess reconstruction accuracy at the mesh level, we report vertex Mean Squared Error (MSE) and L1 Vertex Difference (LVD)~\citep{codetalker} over the full-body SMPL-X mesh. MSE quantifies positional accuracy of reconstructed vertices, while LVD measures the L1 discrepancy of per-vertex velocity, reflecting temporal smoothness:
\begin{equation}
  \text{MSE} = \frac{1}{nT}\sum_{i=1}^{n}\sum_{t=1}^{T}\|\mathbf{f}_{i,t} - \hat{\mathbf{f}}_{i,t}\|^2, \qquad
  \text{LVD} = \frac{1}{n(T-1)}\sum_{i=1}^{n}\sum_{t=2}^{T}\bigl\|\mathbf{v}_{i,t} - \hat{\mathbf{v}}_{i,t}\bigr\|_1,
  \label{eq:mse_lvd}
\end{equation}
where $n$ is the number of mesh vertices, $\mathbf{f}_{i,t}$ and $\hat{\mathbf{f}}_{i,t}$ denote the ground-truth and predicted positions of vertex $i$ at frame $t$, and $\mathbf{v}_{i,t} = \mathbf{f}_{i,t} - \mathbf{f}_{i,t-1}$ is the ground-truth vertex velocity. FaceMSE and FaceLVD apply identical formulas restricted to face mesh vertices, with all body-joint parameters set to zero.

\subsection{Fréchet Gesture Distance (FGD)}
\label{ap:FGD}
FGD~\citep{yoon2020speech} measures the distributional similarity between real and synthesized gesture features via the Fréchet distance:
\begin{equation}
  \text{FGD} = \|\mu_r - \mu_g\|^2 + \operatorname{Tr}\!\left(
    \Sigma_r + \Sigma_g - 2\!\left(\Sigma_r^{1/2}\,\Sigma_g\,\Sigma_r^{1/2}\right)^{1/2}
  \right),
  \label{eq:fgd}
\end{equation}
where $\mu_r, \Sigma_r$ and $\mu_g, \Sigma_g$ are the mean and covariance of the latent feature distributions for ground-truth and generated gestures, respectively. We report FGD under two feature encoders: our RVQ-VAE encoder (concatenating all four part-wise latents) and the publicly available VAESKConv encoder~\citep{liu2024emage}\footnote{\url{https://huggingface.co/H-Liu1997/emage_evaltools}}.

\subsection{Diversity}
\label{ap:Diversity}
Following~\citep{audio2gestures}, Diversity quantifies motion variation across generated clips by computing the average pairwise L1 distance between joint positions of $N$ randomly sampled sequences:
\begin{equation}
  \text{Diversity} = \frac{1}{N(N-1)}\sum_{i=1}^{N}\sum_{j=1}^{N}
    \bigl\|p_i - p_j\bigr\|_1,
  \label{eq:diversity}
\end{equation}
where $p_i$ denotes the joint positions of the $i$-th sampled motion clip with global translation removed. Higher values indicate richer motion dynamics.

\subsection{Beat Constancy (BC) and Normalized Beat Constancy (NBC)}
\label{ap:BC and NBC}
BC~\citep{li2021ai} evaluates speech-motion synchronization by measuring how closely the motion beats of upper-body joints align with audio onset beats. Motion beats are identified as local velocity minima, with velocities normalized by the mean motion amplitude of the test set. For each gesture beat $b_g$, the alignment score is computed as the Gaussian-weighted distance to its nearest audio beat:
\begin{equation}
  \text{BC} = \frac{1}{|g|}\sum_{b_g \in g}
    \exp\!\left(-\frac{\min_{b_a \in a}\,\|b_g - b_a\|^2}{2\sigma^2}\right),
  \label{eq:bc}
\end{equation}
where $g$ and $a$ denote the sets of gesture and audio beats, and $\sigma$ is the Gaussian bandwidth following~\citep{li2021ai}. For RVQ-VAE reconstruction evaluation, we additionally report Normalized Beat Constancy $\text{NBC} = \text{BC}_{\text{rec}} / \text{BC}_{\text{GT}}$, which expresses the reconstructed beat alignment relative to that of the ground-truth sequence.

%% file: sections/appendix/user_study.tex
\subsection{Overview}
\label{ap:slt-overview}
\begin{figure*}[t]
\vspace{-2mm}
  \centering
  \includegraphics[height=13cm]{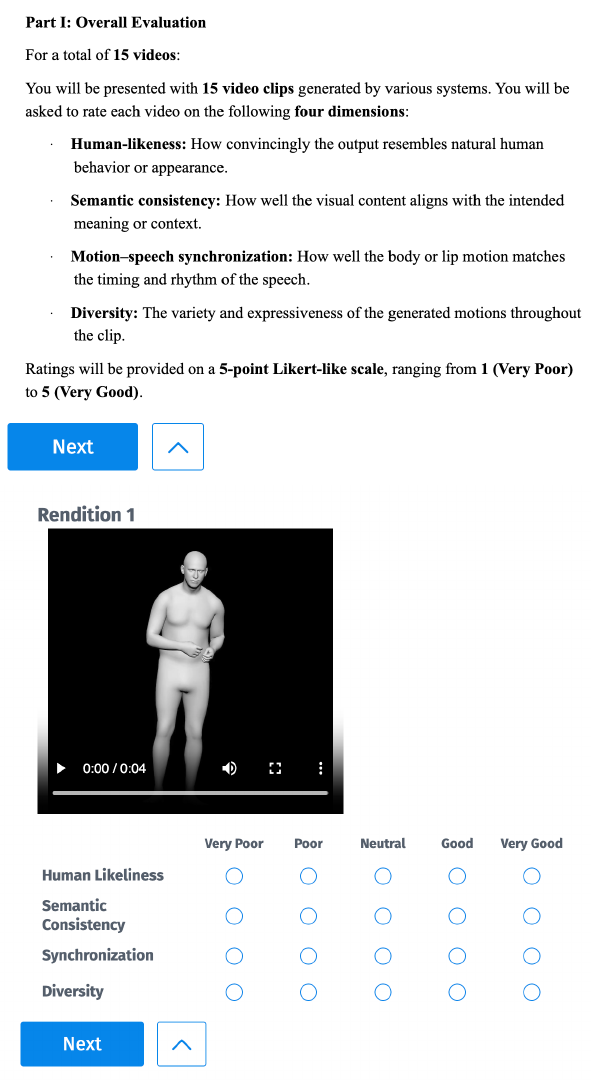}
  \hfill
  \includegraphics[height=13cm]{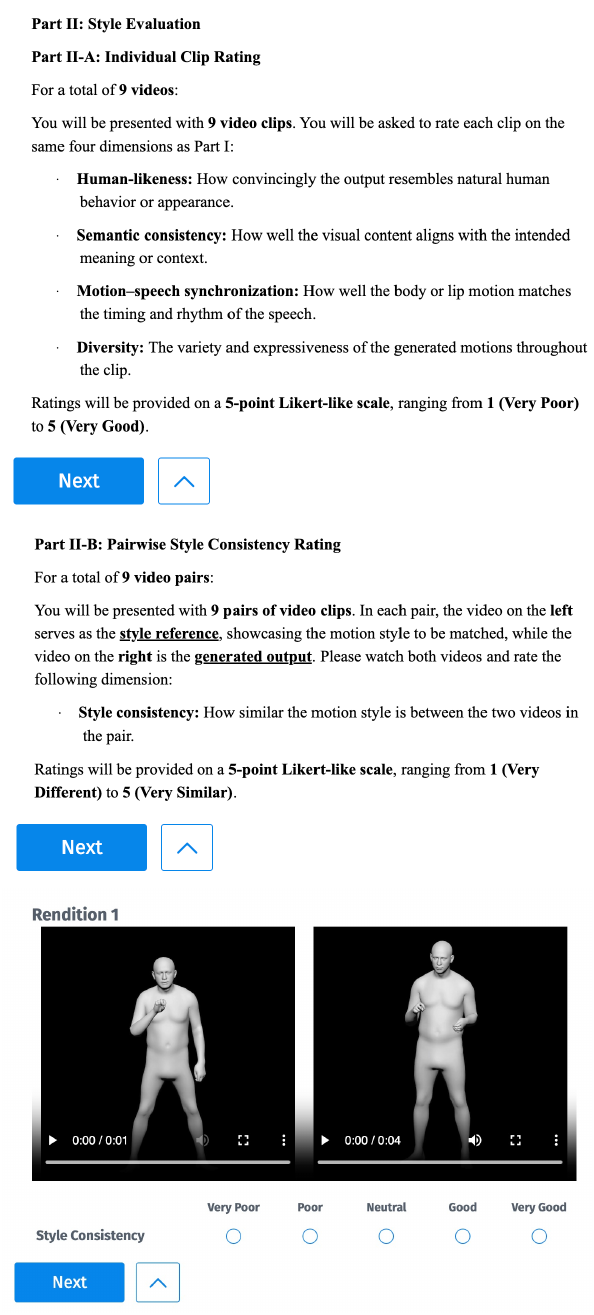}
  \caption{Screenshots of survey pages and instructions presented to participants. (a) Part I: Overall gesture evaluation. (b) Part II: Style-conditioned gesture evaluation.}
  \label{fig:survey}
  \vspace{-3mm}
\end{figure*}

We conduct our subjective evaluation using an online survey hosted on QuestionPro\footnote{\url{https://www.questionpro.com/}}, structured into two parts: (1) overall quality evaluation of generated video clips, and (2) motion style evaluation including both individual clip rating and pairwise style consistency assessment. Each participant completes both parts, with an average completion time of approximately 20--25 minutes. Figure \ref{fig:survey} shows sample survey pages along with participant instructions. Detailed descriptions of the survey structure are provided below.

\subsection{Survey Structure}
\label{ap:slt-Survey Structure}

\paragraph{Part I: Overall Evaluation.}
Participants are presented with 15 video clips generated by various systems. Each clip is rated along the following four dimensions:
\begin{itemize}
    \item \textbf{Human-likeness:} How convincingly the output resembles natural human behavior or appearance.
    \item \textbf{Semantic consistency:} How well the visual content aligns with the intended meaning or context.
    \item \textbf{Motion--speech synchronization:} How well the body motion matches the timing and rhythm of the speech.
    \item \textbf{Diversity:} The variety and expressiveness of the generated motions throughout the clip.
\end{itemize}
All ratings are provided on a 5-point Likert-like scale ranging from 1 (Very Poor) to 5 (Very Good).

\paragraph{Part II: Style Evaluation.}
This part consists of two sections designed to assess motion style quality from complementary perspectives.

\subparagraph{Part II-A: Individual Clip Rating.}
Participants are presented with 9 video clips and asked to rate each on the same four dimensions as Part I.

\subparagraph{Part II-B: Pairwise Style Consistency Rating.}
Participants are presented with 9 video pairs. In each pair, the left video serves as the style reference and the right video is the generated output. Participants rate the following dimension:
\begin{itemize}
    \item \textbf{Style consistency:} How similar the motion style is between the two videos in the pair.
\end{itemize}
Ratings are provided on a 5-point Likert-like scale ranging from 1 (Very Different) to 5 (Very Similar).

\subsection{Additional Notes}
\label{ap:slt-Additional Notes}
\begin{itemize}
    \item Participants are instructed to evaluate motion quality and style rather than personal preference or video fidelity.
    \item All model outputs are anonymized and presented in randomized order to minimize potential bias.
    \item Participants are encouraged to watch each clip in full before submitting ratings.
    \item Participants are advised to view the clips in a quiet environment, preferably with headphones, to ensure accurate perception of speech.
    \item No personal data is collected.
\end{itemize}

%% file: sections/appendix/sup_results.tex
This section provides extended quantitative results complementing the main paper. We first report full-metric evaluations on seen speakers (held-out test sequences from the 20 training speakers) for the VQ-based motion representation comparison, the co-speech gesture generation benchmark, and the style-conditioned gesture generation comparison, including metrics omitted from the main paper due to space constraints. We further provide ablation results on content-style disentanglement quality and a hyperparameter sensitivity analysis for the semantic-aware remasking weights $\alpha$ and $\beta$ in the Content Masked Transformer.

\subsection{VQ-based Motion Representation}
\label{ap:VQ-based Motion Representation}
Table~\ref{tab:vq_seen} reports the seen speaker results for the VQ-based motion representation comparison, extending the zero-shot results presented in the main paper. PersonaGest achieves the best performance on Hands, Lower, JRMSE, FGD, FGD$_\text{sk}$, and NBC, confirming that the gains observed under zero-shot conditions are consistent across both settings. RVQ-VAE (B) remains competitive on MSE and LVD, reflecting its stronger per-vertex reconstruction fidelity, while PersonaGest maintains a clear advantage on distribution quality metrics. APVQ-VAE leads on Face reconstruction but at the cost of substantially worse performance on body joints and distribution metrics.

{
\setlength{\heavyrulewidth}{1.5pt}
\setlength{\lightrulewidth}{0.45pt}
\begin{table*}[t]
\centering
\caption{Quantitative comparison of VQ-based motion representation models under seen speaker settings. For clarity, we report Face$\times10^{-3}$, Upper$\times10^{-3}$, Hands$\times10^{-2}$, Lower$\times10^{-1}$, JRMSE$\times10^{-2}$, MSE$\times10^{-5}$, and LVD$\times10^{-2}$, FGD$\times 10^{-3}$. \textbf{Bold}: best; \underline{underline}: second best.}
\label{tab:vq_seen}
\resizebox{\textwidth}{!}{%
\begin{tabular}{l ccccc cc cccc}
\toprule
\textbf{Model}
  & \textbf{Face $\downarrow$}
  & \textbf{Upper $\downarrow$}
  & \textbf{Hands $\downarrow$}
  & \textbf{Lower $\downarrow$}
  & \textbf{JRMSE $\downarrow$}
  & \textbf{MSE $\downarrow$}
  & \textbf{LVD $\downarrow$}
  & \textbf{FGD $\downarrow$}
  & \textbf{FGD$_\text{sk}$ $\downarrow$}
  & \textbf{NBC $\downarrow$}
  & \textbf{Diversity $\uparrow$} \\
\midrule
VQ-VAE
  & 1.004\scriptsize{±0.335} & 1.230\scriptsize{±1.080} & 2.087\scriptsize{±1.373} & 8.820\scriptsize{±6.610} & 1.282\scriptsize{±0.780}
  & 3.670\scriptsize{±2.070} & 3.530\scriptsize{±1.050}
  & 7.011 & 8.331 & 1.613\scriptsize{±0.556} & 5.624 \\
APVQ-VAE
  & \textbf{0.201}\scriptsize{±0.075} & 0.665\scriptsize{±0.171} & 1.020\scriptsize{±0.833} & 3.970\scriptsize{±3.650} & 0.627\scriptsize{±0.553}
  & 1.640\scriptsize{±0.949} & 2.450\scriptsize{±0.665}
  & 1.212 & 1.697 & 1.803\scriptsize{±0.314} & 7.912 \\
RVQ-VAE (S)
  & 0.543\scriptsize{±0.371} & \underline{0.380}\scriptsize{±0.167} & \underline{0.366}\scriptsize{±0.341} & \underline{0.940}\scriptsize{±0.129} & \underline{0.244}\scriptsize{±0.037}
  & \underline{0.431}\scriptsize{±0.309} & \underline{1.260}\scriptsize{±0.392}
  & 0.395 & 0.753 & 1.294\scriptsize{±0.462} & \textbf{8.761} \\
RVQ-VAE (B)
  & \underline{0.475}\scriptsize{±0.237} & \textbf{0.362}\scriptsize{±0.167} & 0.376\scriptsize{±0.358} & 1.080\scriptsize{±0.180} & 0.248\scriptsize{±0.037}
  & \textbf{0.394}\scriptsize{±0.272} & \textbf{1.220}\scriptsize{±0.361}
  & \underline{0.260} & \underline{0.581} & \underline{1.053}\scriptsize{±0.485} & \underline{8.443} \\
\textbf{PersonaGest}
  & 0.517\scriptsize{±0.365} & 0.414\scriptsize{±0.168} & \textbf{0.286}\scriptsize{±0.280} & \textbf{0.920}\scriptsize{±0.121} & \textbf{0.227}\scriptsize{±0.036}
  & 0.552\scriptsize{±0.084} & 1.280\scriptsize{±0.683}
  & \textbf{0.179} & \textbf{0.483} & \textbf{1.024}\scriptsize{±0.484} & 8.302 \\
\midrule
\textit{p-value}
  & <0.0001 & <0.0001 & <0.0001 & <0.0001 & <0.0001
  & <0.0001 & <0.0001
  & <0.0001 & <0.0001 & <0.0001 & <0.0001 \\
\bottomrule
\end{tabular}%
}
\end{table*}
}

\subsection{Co-speech Gesture Generation Benchmark}
\label{ap:Co-speech Gesture Generation Benchmark}
We evaluate PersonaGest against state-of-the-art co-speech gesture generation methods, including EMAGE~\citep{liu2024emage}\footnote{\url{https://github.com/PantoMatrix/PantoMatrix/}}, MambaTalk~\citep{xu2024mambatalk}\footnote{\url{https://github.com/kkakkkka/MambaTalk}}, EchoMask~\citep{zhang2025echomask}\footnote{\url{https://github.com/Human3DAIGC/EchoMask}}, SemTalk~\citep{zhang2025semtalk}\footnote{\url{https://github.com/Xiangyue-Zhang/SemTalk}}, PyraMotion~\citep{yinpyramotion}\footnote{\url{https://github.com/Williamy946/PyraMotion}}, and GestureLSM~\citep{liu2025gesturelsm}\footnote{\url{https://github.com/andypinxinliu/GestureLSM}}. Table~\ref{tab:unconditional_seen} extends the main paper results with FaceMSE and FaceLVD metrics for the seen speaker setting. PersonaGest achieves the best FGD and FGD$_\text{sk}$, indicating the closest motion distribution to the ground truth. On facial metrics, PersonaGest ranks second behind PyraMotion on FaceMSE but achieves the second-best FaceLVD, demonstrating competitive facial motion quality. GestureLSM does not report facial parameters and is excluded from facial metric ranking.

{
\setlength{\heavyrulewidth}{1.5pt}
\setlength{\lightrulewidth}{0.45pt}
\begin{table*}[t]
\centering
\caption{Quantitative comparison with state-of-the-art co-speech gesture generation models under seen speaker settings. For clarity, we report FGD$\times10^{-5}$, FGD$_\text{sk}$$\times 10^{-1}$, MSE$\times10^{-6}$, LVD$\times10^{-2}$, FaceMSE$\times10^{-8}$, and FaceLVD$\times10^{-5}$. \textbf{Bold}: best; \underline{underline}: second best. $^\dagger$GestureLSM does not generate facial parameters; `---' indicates metric not available.}
\label{tab:unconditional_seen}
\resizebox{\textwidth}{!}{%
\begin{tabular}{l cccc cccc}
\toprule
\textbf{Model}
  & \textbf{FGD $\downarrow$}
  & \textbf{FGD$_\text{sk}$ $\downarrow$}
  & \textbf{BC $\uparrow$}
  & \textbf{Diversity $\uparrow$}
  & \textbf{MSE $\downarrow$}
  & \textbf{LVD $\downarrow$}
  & \textbf{FaceMSE $\downarrow$}
  & \textbf{FaceLVD $\downarrow$} \\
\midrule
EMAGE~(CVPR'24)
  & 3.685 & 2.831 & 0.838\scriptsize{±0.026} & 10.980
  & 1.070\scriptsize{±0.300} & 6.110\scriptsize{±0.860} & 7.820\scriptsize{±2.600} & 8.530\scriptsize{±1.400} \\
MambaTalk~(NeurIPS'24)
  & 3.704 & 2.644 & \textbf{0.884}\scriptsize{±0.020} & 10.585
  & 0.920\scriptsize{±0.270} & 5.910\scriptsize{±0.810} & 7.590\scriptsize{±2.500} & 8.320\scriptsize{±1.400} \\
EchoMask~(MM'25)
  & 3.184 & 3.284 & 0.837\scriptsize{±0.031} & \textbf{12.676}
  & 1.700\scriptsize{±0.520} & 8.210\scriptsize{±1.400} & 7.960\scriptsize{±2.900} & 8.660\scriptsize{±1.600} \\
SemTalk~(ICCV'25)
  & 3.505 & 2.625 & 0.788\scriptsize{±0.027} & \underline{11.960}
  & 1.130\scriptsize{±0.310} & 6.620\scriptsize{±0.860} & 8.290\scriptsize{±3.000} & 8.800\scriptsize{±1.600} \\
PyraMotion~(NeurIPS'25)
  & \underline{2.503} & \underline{1.868} & 0.690\scriptsize{±0.131} & 5.599
  & \underline{0.770}\scriptsize{±0.620} & \underline{4.570}\scriptsize{±2.040} & \textbf{3.810}\scriptsize{±1.900} & \textbf{5.680}\scriptsize{±1.400} \\
GestureLSM$^\dagger$~(ICCV'25)
  & 2.936 & 2.818 & 0.673\scriptsize{±0.044} & 8.450
  & \textbf{0.760}\scriptsize{±0.250} & 5.340\scriptsize{±0.790} & --- & --- \\
\textbf{PersonaGest~(Ours)}
  & \textbf{2.475} & \textbf{1.414} & \underline{0.859}\scriptsize{±0.091} & 11.053
  & 0.800\scriptsize{±0.370} & \textbf{4.290}\scriptsize{±1.230} & \underline{5.310}\scriptsize{±1.700} & \underline{6.850}\scriptsize{±1.000} \\
\midrule
\textit{p-value}
  & <0.0001 & <0.0001 & <0.0001 & <0.05
  & <0.0001 & <0.0001 & <0.0001 & <0.0001 \\
\bottomrule
\end{tabular}%
}
\end{table*}
}

\subsection{Style-conditioned Co-speech Gesture Generation}
\label{ap:Style-conditioned Co-speech Gesture Generation}
{
\setlength{\heavyrulewidth}{1.5pt}
\setlength{\lightrulewidth}{0.45pt}
\begin{table*}[t!]
\centering
\caption{Quantitative comparison with style-conditioned co-speech gesture generation models under seen speaker settings. For clarity, we report FGD$\times10^{-5}$, FGD$_\text{sk}$$\times 10^{-1}$, MSE$\times10^{-5}$, LVD$\times10^{-2}$, FaceMSE$\times10^{-9}$, and FaceLVD$\times10^{-6}$. \textbf{Bold}: best; \underline{underline}: second best. $^\dagger$SynTalker does not generate facial parameters; `---' indicates metric not available.}
\label{tab:conditional_seen}
\resizebox{\textwidth}{!}{%
\begin{tabular}{l cccc cc cc}
\toprule
\textbf{Model}
  & \textbf{FGD $\downarrow$}
  & \textbf{FGD$_\text{sk}$ $\downarrow$}
  & \textbf{BC $\uparrow$}
  & \textbf{Diversity $\uparrow$}
  & \textbf{MSE $\downarrow$}
  & \textbf{LVD $\downarrow$}
  & \textbf{FaceMSE $\downarrow$}
  & \textbf{FaceLVD $\downarrow$} \\
\midrule
SynTalker$^\dagger$
  & 3.293 & 2.101 & \underline{0.605}\scriptsize{±0.087} & \textbf{10.448}
  & 8.500\scriptsize{±2.900} & 5.630\scriptsize{±0.850} & --- & --- \\
ZeroEGGS
  & \underline{2.871} & \underline{1.971} & 0.587\scriptsize{±0.015} & 2.541
  & \textbf{4.200}\scriptsize{±2.700} & \textbf{3.280}\scriptsize{±1.090} & \underline{1.180}\scriptsize{±0.470} & \underline{1.130}\scriptsize{±0.220} \\
\textbf{PersonaGest~(Ours)}
  & \textbf{2.480} & \textbf{1.462} & \textbf{0.629}\scriptsize{±0.091} & \underline{6.053}
  & \underline{8.120}\scriptsize{±3.600} & \underline{4.980}\scriptsize{±0.970} & \textbf{0.545}\scriptsize{±0.180} & \textbf{0.693}\scriptsize{±0.120} \\
\midrule
\textit{p-value}
  & <0.0001 & <0.0001 & <0.0001 & <0.0001
  & <0.01 & <0.0001 & <0.0001 & <0.0001 \\
\bottomrule
\end{tabular}%
}
\end{table*}
}

We compare PersonaGest with two style-conditioned co-speech generation models SynTalker~\citep{chen2024enabling}\footnote{\url{https://github.com/RobinWitch/SynTalker}} and ZeroEGGS~\citep{ghorbani2023zeroeggs}\footnote{\url{https://github.com/ubisoft/ubisoft-laforge-ZeroEGGS}}, that support motion style prompts.
Table~\ref{tab:conditional_seen} extends the main paper results with FaceMSE and FaceLVD for the seen speaker setting. PersonaGest achieves the best FaceMSE and FaceLVD by a clear margin, demonstrating that style conditioning does not compromise facial reconstruction quality. ZeroEGGS achieves the best body MSE and LVD but with significantly lower Diversity, consistent with our main paper analysis. SynTalker does not generate facial parameters and is excluded from facial metric ranking.

\subsection{Semantic Fidelity Analysis}
\label{ap:Semantic Fidelity Analysis}
To evaluate whether the generated gestures preserve speech-aligned semantics, we train a gesture category classifier on ground-truth motion from the BEAT2 training set. Following the 1D CNN+LSTM architecture in BEAT~\citep{liu2022beat}, the classifier takes per-frame pose sequences as input and predicts one of five gesture categories defined in BEAT2~\citep{liu2024emage} (nogesture, beat, deictic, iconic, metaphoric), with focal loss~\citep{lin2017focal} to address the severe class imbalance inherent in naturalistic gesture data. The trained classifier is then applied to both GT and generated motion on the test set to measure semantic preservation. As shown in Table~\ref{tab:semantic}, PersonaGest closely approaches the GT upper bound across all metrics with a gap of less than 4\%, demonstrating that the generated gestures preserve meaningful semantic categories aligned with speech content.

{
\setlength{\heavyrulewidth}{1.5pt}
\setlength{\lightrulewidth}{0.45pt}
\begin{table}[t!]
\small
\centering
\caption{Gesture semantic classification results on GT and generated motions.}
\label{tab:semantic}
\vspace{4pt}
\resizebox{0.45\columnwidth}{!}{%
\begin{tabular}{lcccc}
\toprule
& Acc. & Precision & Recall & F1 \\
\midrule
GT         & 81.49 & 79.24 & 81.49 & 80.26 \\
\textbf{Ours} & 77.67 & 75.26 & 78.85 & 77.01 \\
\bottomrule
\end{tabular}
}
\end{table}
}

\subsection{Ablation on Content-Style Disentanglement}
\label{ap:Ablation on Content-Style Disentanglement}
{
\setlength{\heavyrulewidth}{1.5pt}
\setlength{\lightrulewidth}{0.45pt}
\begin{table}[t!]
\centering
\caption{Ablation on content-style disentanglement training objectives. FGD(20)/FGD$_\text{sk}$(20): seen speakers; FGD(5)/FGD$_\text{sk}$(5): zero-shot speakers. \textbf{Bold}: best.}
\label{tab:ablation_dis}
\vspace{7pt}
\resizebox{0.6\columnwidth}{!}{%
\begin{tabular}{l cccc}
\toprule
\textbf{Variant} & \textbf{FGD(20)}$\downarrow$ & \textbf{FGD$_\text{sk}$(20)}$\downarrow$ & \textbf{FGD(5)}$\downarrow$ & \textbf{FGD$_\text{sk}$(5)}$\downarrow$ \\
\midrule
-w/o $\mathcal{L}_{\rm cl}$ & 2.231 & 1.029 & 1.349 & 0.740 \\
-w/o $\mathcal{L}_{\rm cl}$ + $\mathcal{L}_{\rm phone}$ & 3.255 & 1.035 & 1.467 & 0.871 \\
\textbf{Ours} & \textbf{1.787} & \textbf{0.483} & \textbf{0.956} & \textbf{0.517} \\
\bottomrule
\end{tabular}
}
\end{table}
}

\begin{figure}[t!]
  \centering
  \includegraphics[width=0.49\linewidth]{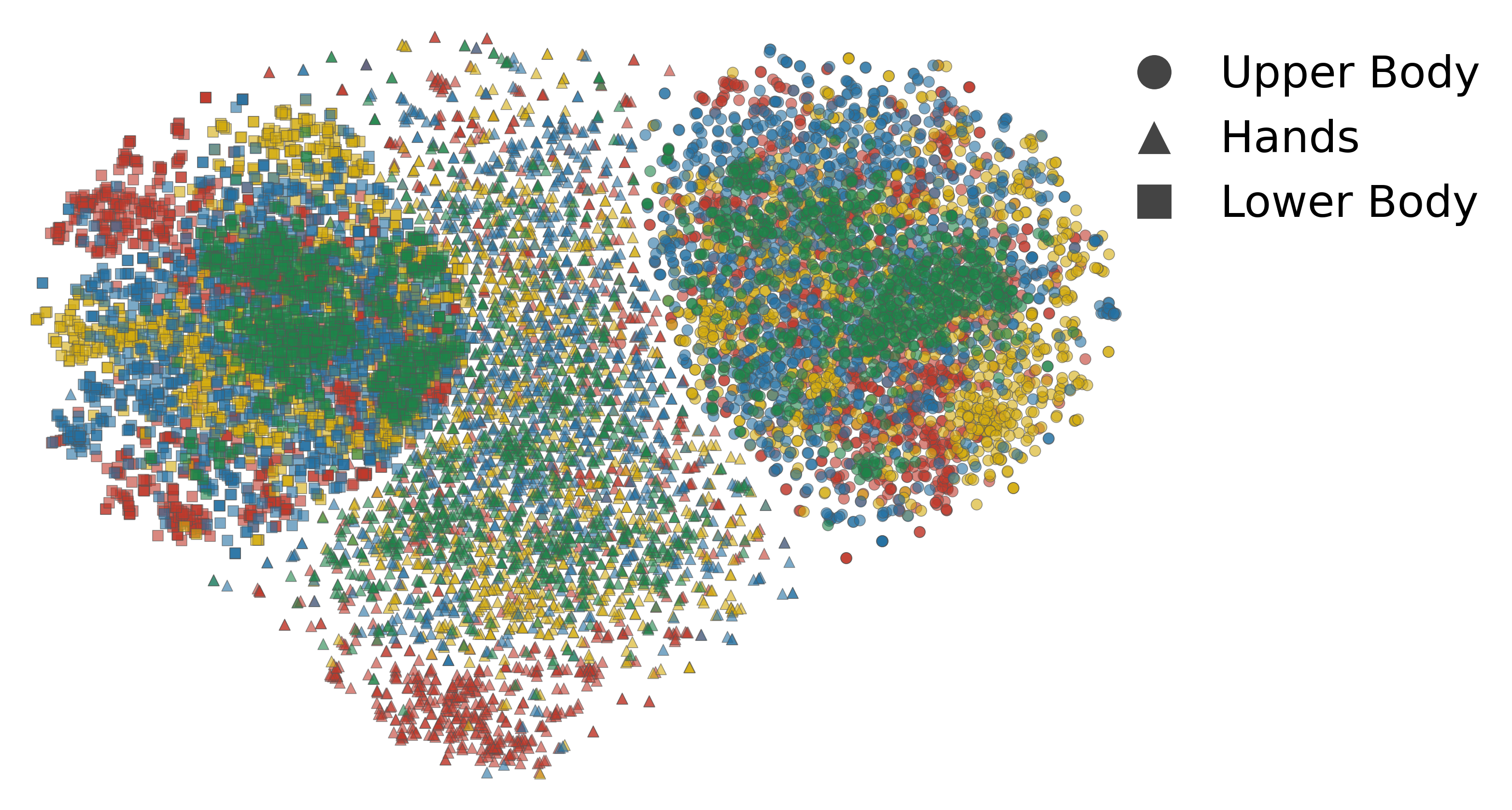}
  \hfill
  \includegraphics[width=0.49\linewidth]{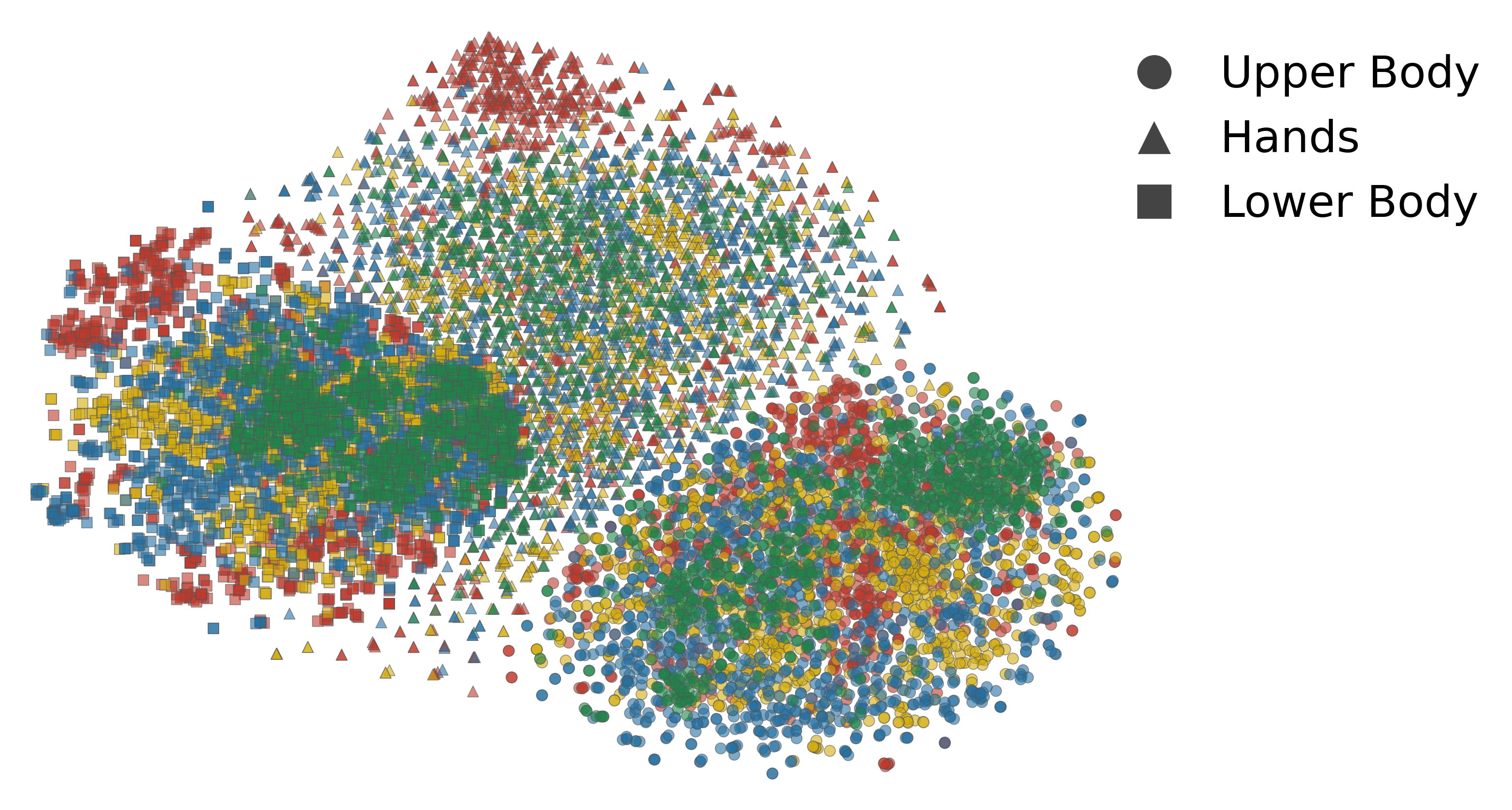}
  \caption{T-SNE visualization of style embeddings for ablation variants, colored by speaker identity. Left: -w/o $\mathcal{L}_{\rm cl}$ + $\mathcal{L}_{\rm phone}$. Right: -w/o $\mathcal{L}_{\rm cl}$.}
  \label{fig:tsne_ablation}
\end{figure}

Beyond the speaker identification experiments, we further validate disentanglement through a style transition experiment following~\citep{zargarbashi2026vq}: given two motion prompts, content tokens are extracted from prompt A while style tokens are extracted from prompt B, and the two are combined and decoded by the RVQ-VAE to reconstruct the final motion. FGD is computed against the reference speaker's motion to measure style fidelity. As shown in Table~\ref{tab:ablation_dis}, removing only $\mathcal{L}_{\rm cl}$ already leads to substantially worse FGD on both seen and zero-shot speakers, and jointly removing $\mathcal{L}_{\rm cl}$ and $\mathcal{L}_{\rm phone}$ causes the largest degradation across all metrics. These results confirm that both the contrastive objective and phoneme supervision are essential for learning disentangled style representations that generalize to unseen speakers. As shown in Figure~\ref{fig:tsne_ablation}, both ablation variants produce heavily intermixed style embeddings without speaker clusters, confirming that $\mathcal{L}_{\rm cl}$ is essential to capture speaker-specific motion characteristics. 

\subsection{Ablation on Semantic-Aware Remasking}
\label{ap:Ablation on Semantic-Aware Remasking}
{
\setlength{\heavyrulewidth}{1.5pt}
\setlength{\lightrulewidth}{0.45pt}
\begin{table}[t!]
\centering
\caption{Ablation on semantic-aware remasking hyperparameters,
evaluated on zero-shot unseen speakers. \colorbox{gray!15}{Shaded}
rows indicate our final configuration. \textbf{Bold} denotes the
best.}
\label{tab:ablation_remasking}
\setlength{\tabcolsep}{4pt}
\renewcommand{\arraystretch}{0.85}
\small
\begin{subtable}[t]{0.42\linewidth}
\centering
\caption{Remask weighting $(\alpha,\,\beta)$}
\begin{tabular}{lrrr}
\toprule
$(\alpha,\,\beta)$
  & \textbf{FGD}$\downarrow$
  & \textbf{Div}$\uparrow$
  & \textbf{LVD}$\downarrow$ \\
\midrule
$(0,\,1)$     & 2.463 & 11.250 & 4.72 \\
$(1,\,0)$     & 2.642 & 11.054 & 5.19 \\
$(0.5,\,0.5)$ & 2.353 & 11.748 & 4.57 \\
\rowcolor{gray!15}
$(0.6,\,0.4)$ & \textbf{2.311} & \textbf{11.970} & 4.63 \\
$(0.4,\,0.6)$ & 2.348 & 11.859 & \textbf{4.32} \\
\bottomrule
\end{tabular}
\end{subtable}
\hspace{8pt}
\begin{subtable}[t]{0.42\linewidth}
\centering
\caption{Randomisation scale}
\begin{tabular}{crrr}
\toprule
$r$
  & \textbf{FGD}$\downarrow$
  & \textbf{Div}$\uparrow$
  & \textbf{LVD}$\downarrow$ \\
\midrule
0.0 & 2.355 & 11.365 & \textbf{4.57} \\
\rowcolor{gray!15}
0.2 & \textbf{2.311} & 11.970 & 4.63 \\
0.4 & 2.393 & 12.456 & 4.79 \\
0.6 & 2.421 & 12.779 & 4.82 \\
0.8 & 2.534 & 12.964 & 4.77 \\
1.0 & 2.548 & \textbf{12.982} & 4.95 \\
\bottomrule
\end{tabular}
\end{subtable}
\end{table}
}

We ablate two inference-time hyperparameters of our semantic-aware
remasking strategy on zero-shot unseen speakers
(Table~\ref{tab:ablation_remasking}).
Relying on either signal alone underperforms combined strategies,
with the semantic-only variant yielding the worst FGD; a slight
emphasis on semantic priority $(\alpha{=}0.6, \beta{=}0.4)$
achieves the best FGD and Diversity, confirming that the two cues
are complementary.
For the randomisation scale, deterministic remasking suppresses
diversity, whereas a small perturbation ($r{=}0.2$) improves both
FGD and Diversity; quality then degrades monotonically as $r$
increases, collapsing to random remasking at $r{=}1.0$.
We adopt $r{=}0.2$ as our default.

%% file: sections/appendix/sup_vis.tex
\begin{figure}[t!]
    \centering
    \includegraphics[width=\linewidth]{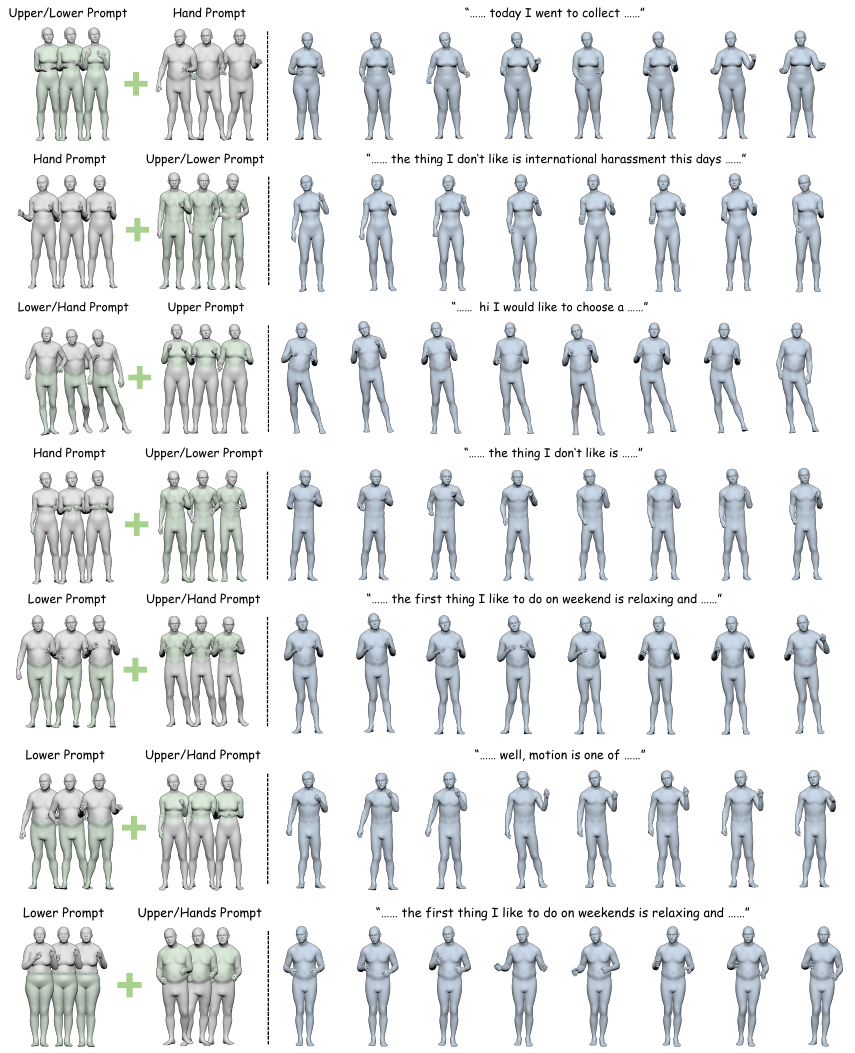}
    \caption{Part-wise style-controlled gesture generation. By tokenising
two motion references and combining tokens from different body parts,
our model generates motion that reflects each reference in its corresponding body region.}
    \label{fig:showcase}
\end{figure}

Figure~\ref{fig:showcase} presents additional qualitative results
demonstrating part-wise style controllability. Each row shows a
generated sequence conditioned on two complementary motion style
references, where upper body, hand, and lower body styles are
independently sourced from different speakers and faithfully
reflected in the output.

%% file: sections/appendix/limit.tex
PersonaGest uses motion examples as style prompts, which provides an intuitive 
and flexible means of style specification but requires the user to supply a 
reference motion clip. Future work could explore alternative style conditioning 
inputs, such as natural language descriptions or emotion-based prompts, to 
further reduce the dependency on reference motion and broaden the applicability 
of the framework. A unified model that supports multiple style specification 
modalities would enable more versatile and accessible gesture generation in 
real-world applications.

Additionally, our evaluation is conducted solely on the  English-language co-speech dataset, which reflects a broader limitation of the field: publicly available co-speech gesture datasets remain scarce, and existing benchmarks are largely restricted to a single language and cultural context. Gesture styles and their relationship to speech may vary across languages and cultures, and the learned style representations may not generalize beyond the distribution seen during training. Future work could address this by constructing more diverse, multilingual co-speech datasets, and by extending PersonaGest to support cross-lingual and cross-cultural style transfer, where a speaker's gestural style is conditioned on references from a different linguistic or cultural background. 

%% file: sections/appendix/broader_impact.tex
This work presents a framework for co-speech gesture generation that synthesises full-body motion from speech audio, conditioning on a short motion reference to capture a speaker's individual gestural style. Style-consistent gesture generation has broad potential to improve human-computer interaction across a range of domains. In virtual avatar and digital human applications, the ability to produce identity-consistent non-verbal behaviour is directly relevant to telepresence, virtual meetings, and social virtual reality, where the absence of personalised body language remains a barrier to authentic communication. In the entertainment and creative industries, style-driven gesture synthesis can substantially reduce the time and cost associated with professional motion capture, lowering the barrier to entry for independent creators and smaller studios. In education and training, virtual instructors with style-consistent gestural behaviour can enhance learner engagement, consistent with established findings that congruent gestures improve speech comprehension.

As with any model that conditions generation on a motion style reference, there is a potential for misuse in producing synthetic content that imitates a specific individual's gestural behaviour without their consent. The ability to transfer style from a short motion clip introduces particular concerns around identity misrepresentation in public-facing applications. We encourage responsible deployment practices, including appropriate disclosure of synthetic content, and advocate for future work on robust detection methods for style-transferred human motion.